\documentclass[12pt,reprint,amsmath,amssymb,aps,prd,twocolumn,superscriptaddress,floatfix,nofootinbib,longbibliography]{revtex4-2}

\usepackage{mathtools}
\usepackage{braket}

\usepackage{tikz}
\usepackage[colorlinks, allcolors=blue]{hyperref}
\usepackage{booktabs}
\usepackage{multirow}

\usepackage{csquotes}
\usepackage{amsthm}

\newcommand{\Z}{{\mathbb{Z}}}
\newcommand{\R}{{\mathbb{R}}}
\newcommand{\C}{{\mathbb{C}}}

\newcommand{\pqty}[1]{\left( #1 \right)}
\newcommand{\bqty}[1]{\left[ #1 \right]}
\newcommand{\abs}[1]{\left\lvert #1\right\rvert}
\newcommand{\norm}[1]{\left\lVert #1\right\rVert}
\newcommand{\expval}[1]{\langle #1 \rangle}

\newcommand {\dv}[3][ ]{
  \ifx #1 { }
    \frac{d #2}{d #3}
  \else
    \frac{d^{#1} #2}{d #3^{#1}}
  \fi
}

\newcommand {\pdv}[3][ ]{
  \ifx #1 { }
    \frac{\partial #2}{\partial #3}
  \else
    \frac{\partial^{#1} #2}{\partial #3^{#1}}
  \fi
}

\renewcommand{\Im}{\operatorname{Im}}

\newcommand{\SU}{\mathrm{SU}}
\newcommand{\U}{\mathrm{U}}

\begin{abstract}

Recently, a version of $\U(1)$ lattice gauge theory on the cubic lattice with a monopole enhancement term was considered in the literature. In this work, we further our understanding of this theory with a mixture of analytical and numerical methods. We show that for sufficiently large monopole enhancement, the monopoles are no longer dilute and they arrange themselves into an antiferromagnetic, Néel-ordered ground state. Nonetheless, a variant of Polyakov's argument for the dilute excitations over the dense ground state leads to an effective sine-Gordon model which is qualitatively similar to that of the standard theory. We also predict analytically the presence of first-order transitions as the monopole chemical potential is varied at fixed gauge coupling, corresponding to a change in the dominant monopole charge. Similarly, we predict a transition in the Ising universality class as the gauge coupling is varied. We then use numerical methods to study the phase diagram of the theory and check some of our predictions.

\end{abstract}

\begin{document}

\title{Monopole crystal and effective theory for a monopole-enhanced 3D U(1) gauge theory}

\author{Alessandro Mariani}
\affiliation{Physics Department, University of Turin \& INFN, Turin unit,\\ Via Pietro Giuria 1, I-10125 Turin, Italy}

\maketitle

\newpage
 
\section{Introduction}

The three-dimensional U(1) lattice gauge theory is a key testbed for our understanding of confinement. It is remarkably well-understood analytically, via Polyakov's effective theory \cite{Polyakov1977}. It is confining for all values of the gauge coupling, and its confinement mechanism is well-understood as due to monopoles. Confinement has also been rigorously proven \cite{GopfMack}. The physical degree of freedom is a massive scalar \textquote{photoball} of mass $m$. These analytical predictions have also found numerical confirmation \cite{TepAthen, Loan2003, Loan:2003wy, Caselle:2025_Pressure}. 

A peculiar property of this theory is that, in the continuum limit, the dimensionless ratio of the mass to the string tension is zero,
\begin{equation}
    \frac{m}{\sqrt\sigma} \to 0 \ .
\end{equation}
This is unlike $\SU(N)$ lattice gauge theories in three or four dimensions, where the ratio is instead constant in the continuum limit. Due to this peculiar behavior, one can potentially take different continuum limits depending on which quantity is held fixed. We will elaborate on this further in later sections.  This also means that there is a regime where $m^2 \ll \sigma$, which causes the photoball to have a non-trivial interaction with the massless transverse modes of the confining string, leading to a non-universal form of the confining string potential. This was also predicted analytically \cite{Aharony:2024ctf} and checked numerically \cite{Caselle:2025_String}. 

It is interesting to consider whether a deformation of the theory might lead to a continuum limit where the mass to string tension ratio behaves more conventionally, i.e. is a non-zero constant, 
\begin{equation}
    \frac{m}{\sqrt\sigma} \to \mathrm{const} \neq 0 \ .
\end{equation}
Recently, Athenodorou, Bonati and Soler Calero \cite{athenodorou2026differentkindcontinuumlimit} considered this problem and introduced a monopole-enhancement term (i.e. a chemical potential for the monopoles) in the action. With numerical methods, they then looked for a line of constant physics in parameter space where the mass to string tension ratio stays constant.

In this work, we consider the standard $\U(1)$ lattice gauge theory on the cubic lattice, and add the same monopole-enhancing term as in \cite{athenodorou2026differentkindcontinuumlimit}. We then derive several analytical results for this theory in the large $\beta$ regime. We find that the behavior of the theory is controlled by the ratio $\mu/\beta$ between the monopole chemical potential and the gauge coupling. For sufficiently large $\mu/\beta$, the monopole ground state is a Néel-ordered antiferromagnet which spontaneously breaks a $\Z_2$ charge conjugation symmetry as well as translations by one lattice spacing. In this sense, the monopoles form a \textquote{crystal}. The dominant charge of the monopoles turns out to be determined by the $\mu/\beta$ ratio and as the ratio is varied, one finds first-order transitions between Néel-ordered ground states with different dominant charges. By considering the excitations on top of the dense monopole background and integrating out short-distance fluctuations, one then obtains a version of Polyakov's effective theory for the model with $\mu \neq 0$. This turns out to be a sine-Gordon model, qualitatively the same as for the standard theory. In particular, we show that the ratio $m/\sqrt{\sigma}\to 0$ also for $\mu \neq 0$. Crystalline confined phases had previously been observed in other models of gauge theories, where however the charges are electric rather than magnetic \cite{QuantumLink, QuantumDimer1, QuantumDimer2, QuantumSpinLadder1, QuantumSpinLadder2, Staggered}. One of these models \cite{Staggered} turned out to be related to a limit of $\SU(2)$ Super-Yang-Mills theory \cite{Bub:2020mff}.

We then check some of our predictions for the monopole distribution numerically. In particular, we confirm that the system undergoes first-order transitions as $\mu/\beta$ is varied at fixed large enough $\beta$. We then consider whether the system might undergo a transition (which we predict would be in the Ising universality class) and find a phase transition as $\beta$ is varied at fixed $\mu/\beta$, between a disordered phase and a dense monopole crystal phase. We find the location of the transition and find some confirmation of Ising universality. 

\section{3D U(1) lattice gauge theory with and without monopole enhancement}
\label{sec:the theory}

In the standard path-integral representation of $\U(1)$ lattice gauge theory, we assign to each link $l$ of the cubic lattice a $\U(1)$-valued variable $\varphi_l$. In this work, we adopt the Villain formulation of the theory \cite{Villain:1974ir, Berezinskii:1970pzv}, in which case we also have an integer-valued field $m_p \in \Z$ associated with each lattice plaquette $p$. The partition function is given by
\begin{multline}
    \label{eq:u(1) partition function}
    Z = \pqty{\prod_{l \in \mathrm{links}} \int_\R d\varphi_l} \pqty{\prod_{p \in \mathrm{plaq}} \sum_{m_p \in \Z}} \times \\
    \times  \exp{\bqty{ -\frac{\beta}{2}\sum_p ((d\varphi)_p + 2\pi m_p)^2+\mu \sum_{c} \abs{dm_c} }} \ ,
\end{multline}
where $l$ runs over links connecting neighboring sites of the lattice, $p$ runs over the plaquettes of the lattice and $c$ over lattice cubes. The plaquette variable $d\varphi_p$ associated to plaquette $p$ is given by the ordered sum of the four links (labelled from $1$ to $4$) in the plaquette,
\begin{equation} \label{eq:plaquette}
(d\varphi)_p \equiv \varphi_1 + \varphi_2 - \varphi_3 - \varphi_4 \ .
\end{equation}
Similarly, $dm_c$ is the ordered sum of the six plaquette variables $m_p$ associated to each lattice cube $c$. Apart from the fact that we use the Villain action instead of the Wilson action, our conventions are the same as those of \cite{athenodorou2026differentkindcontinuumlimit}. In the above equations we have used the language of lattice differential forms, and we will make extensive use of this in what follows. We refer the reader to Appendix \ref{sec:lattice differential forms} for an introduction to this topic. 

In the partition function eq.\eqref{eq:u(1) partition function}, we have added to the action a monopole chemical potential $\sim -\mu \abs{dm_c}$ following \cite{athenodorou2026differentkindcontinuumlimit}. If $\mu < 0$, then the monopoles cost extra energy and are therefore suppressed. If instead $\mu > 0$, then the monopoles are enhanced as their energy cost is lowered. The standard theory corresponds to $\mu=0$. We will give a more thorough discussion of the detection and effect of monopoles in the next section.

Note that we have chosen the gauge field in eq.\eqref{eq:u(1) partition function} to be non-compact, i.e. we integrate over all real numbers rather than only over $[0,2\pi)$. This allows some minor simplifications in a later derivation and in any case the two choices differ by a constant infinite factor which cancels in any expectation value. The equivalence of the two constructions is also clear from seeing the $\U(1)$ gauge symmetry of the Villain action as arising from gauging (via the introduction of the integer field) a $\Z$ subgroup of an $\R$-symmetric theory, leading to $\R/\Z \cong \U(1)$. The periodicity of the $\varphi$ field can also be seen explicitly, i.e. we can perform the local transformation
\begin{equation}
    \varphi_l \to \varphi_l -2\pi n_l \ ,
\end{equation}
for arbitrary $n_l$, which leaves the action invariant as long as it is accompanied by an appropriate shift in $m$, i.e.
\begin{equation}
    \label{eq:integer shift Z gauge invariance}
    m_p \to m_p + (dn)_p \ .
\end{equation}
Note that this transformation leaves the monopole number invariant since
\begin{equation}
    dm_c \to dm_c + (d^2n)_c=dm_c \ .
\end{equation}
The action eq.\eqref{eq:u(1) partition function} is also symmetric under standard gauge transformations, i.e. 
\begin{equation}
    \varphi_l \to \varphi_l +d\alpha_l
\end{equation}
where $\alpha_x$ is a collection of site-based real numbers. Then, as usual, one has
\begin{equation}
    d\varphi_p \to d\varphi_p + d^2\alpha_p=d\varphi_p \ ,
\end{equation}
which ensures the symmetry of the action.

\subsection{Monopoles}\label{sec:monopoles}

In three spacetime dimensions, monopoles are point-like topological excitations. In the Villain formulation, monopoles are detected via the auxiliary field $m$, and in particular, one says that there is a monopole inside a lattice cube $c$ if 
\begin{equation}
    dm_c \neq 0 \ .
\end{equation}
The charge of the monopole is then just $dm_c \in \Z$. This mimics the continuum definition of a monopole, whereby one detects the monopole charge enclosed by a surface $S$ via the total flux through the surface (first Chern class)
\begin{equation}
    \frac{1}{2\pi} \int_S F \in \Z \ .
\end{equation}
Under the identification $F \sim d\varphi_p + 2\pi m_p$, which is natural looking at the Villain action, one sees that the total flux through the cube is just $dm_c$ owing to cancellations in the term involving plaquette variables. A similar definition can be given for the Wilson action, as we shall see below. Note that, since $dm$ is a total derivative, on a finite periodic box the total monopole charge is zero,
\begin{equation}
    \sum_c dm_c =0 \ .
\end{equation}

In the limit $\mu \to -\infty$, the monopoles are completely suppressed and one has $dm_c=0$ on every cube. In infinite volume, the lattice Poincaré lemma (see Appendix \ref{sec:lattice differential forms}) then implies that there exists an integer-valued field $n_l \in \Z$ living on lattice links, such that
\begin{equation}
    \label{eq:m is exact}
    m = dn \ .
\end{equation}
Then, the partition function takes the form (here we choose to integrate in the range $[0, 2\pi)$),
\begin{multline}
    Z = \pqty{\prod_{l \in \mathrm{links}} \sum_{n_l \in \Z} \int_{0}^{2\pi} d\varphi_l} \times \\
    \times  \exp{\bqty{ -\frac{\beta}{2}\sum_p ((d\varphi)_p + 2\pi dn_p)^2 }} \ .
\end{multline}
One can then change variables to
\begin{equation}
    A_l = \varphi_l +2\pi n_l \ .
\end{equation}
With the new variable, the integration domain is $[2\pi (-n_l), 2\pi (-n_l+1)]$ and the integration domain becomes $(-\infty, +\infty)$ after summing over $n$. One then has
\begin{equation}
    \label{eq:R partition function}
    Z = \pqty{\prod_{l \in \mathrm{links}} \int_\R dA_l} \exp{\bqty{ -\frac{\beta}{2}\sum_p (dA)_p^2 }} \ ,
\end{equation}
which is the partition function of a $\R$-valued gauge field. This theory has a single massless phase (by a change of variable, the partition function is in fact independent of $\beta$). In finite volume, the Poincaré lemma on the torus (see Appendix \ref{sec:lattice differential forms}) allows one to take into account winding corrections, but these can be shown to be exponentially small in the volume and therefore irrelevant for most observables. The fact that removing the monopoles from $\U(1)$ lattice gauge theory leads to $\R$ lattice gauge theory has long been known \cite{Mack1982,Banks:1977cc, Villain:1974ir, GopfMack}. Recently, the theory with the constraint $dm_c=0$ has received attention in the literature and has been dubbed the \textquote{modified Villain model} \cite{Sulejmanpasic:2019ytl, Gorantla:2021svj}.

While in four spacetime dimensions, the continuum limit of the theory exists only in the phase where monopoles are irrelevant and therefore the monopoles can be considered \textquote{lattice artifacts}, in three dimensions this is not the case. The monopoles are relevant in the phase connected to the continuum limit, and are in fact responsible for confinement.

In the continuum, $\U(1)$ lattice gauge theory in three spacetime dimensions is not UV-complete, in the sense that one must decide whether the theory includes monopoles or not. A review of some of these facts can be found in Section 8 of \cite{Tong}. One possible choice is free electrodynamics without monopoles, in which case it is the theory of a massless photon and static charges interact via a Coulomb potential. This corresponds to the lattice gauge theory with gauge group $\R$. On the other hand, one could obtain a $\U(1)$ gauge theory by breaking an $\SU(2)$ symmetry via the Higgs mechanism; in this case the theory has monopoles and it is confining. The effective theory in this case is a sine-Gordon model as for the lattice $\U(1)$ theory, but since it is not renormalizable in three spacetime dimensions it cannot be a target for a continuum limit. In fact the lattice theory's continuum limit is a free massive scalar, and it is not clear what alternative continuum theory could be a target for a lattice regularization.

In the continuum, there also is some debate over the distinction between gauge theories with different gauge
groups but same Lie algebras (such as $\R$ and $\U(1)$, which share the same Lie algebra $\R$) \cite{Harlow:2018tng}. As is clear from the discussion in this section, on the lattice the distinction between $\U(1)$ and $\R$ is in the presence vs absence of monopoles.

\subsection{Wilson vs Villain}
\label{sec:wilson vs villain}

The Villain action is not the only option and one may in fact adopt instead the more commonly used Wilson action, whose partition function reads
\begin{equation}
    Z = \pqty{\prod_{l \in \mathrm{links}} \int_{0}^{2\pi} d\varphi_l} \exp{\bqty{ \beta\sum_p \cos(d\varphi_p) }} \ .
\end{equation}
In particular, the Wilson action was used in the previously cited study which introduced the monopole-enhanced theory \cite{athenodorou2026differentkindcontinuumlimit}. In all cases in the literature, the Wilson and Villain action are found to be equivalent, in the sense that
\begin{enumerate}
    \item They share the same qualitative phase diagram, i.e. the same phases, although non-renormalized quantities such as the location of the critical points may differ between the two. 
    \item At a critical point / continuum limit, they agree on all renormalized quantities. 
\end{enumerate}
In particular, all theoretical predictions for the 3D $\U(1)$ gauge theory have been computed using the Villain action \cite{GopfMack, Aharony:2024ctf} and verified to hold also for the Wilson action \cite{TepAthen, Caselle:2025_String, Caselle:2025_Pressure}.

\begin{table}[]
    \centering
    \begin{tabular}{c|c}
        $\beta_W$ & $\beta_V$\\
        \hline
         1.0 & 0.62 \\
         1.7 & 1.13 \\
         2.0 & 1.39 \\
         2.2 & 1.58 \\
         2.4 & 1.77 \\
         3.0 & 2.37 \\
    \end{tabular}
    \caption{Comparison between Wilson and Villain gauge couplings, following eq.\eqref{eq:wilson villain coupling relation}.}
    \label{tab:wilson villain coupling comparison}
\end{table}

By matching the first Fourier coefficient of the respective Boltzmann weights, one can derive an approximate relation between the couplings for the two theories,
\begin{equation}
    \label{eq:wilson villain coupling relation}
    \beta_V \cong \frac{1}{2 \log(I_0(\beta_W)/I_1(\beta_W))} \ ,
\end{equation}
which is valid for large $\beta$ and where $I_0$ is a modified Bessel function of the first kind. Using asymptotic formulas for the Bessel functions, one sees that for large $\beta$
\begin{equation}
    \beta_V \approx \beta_W -\frac12 + \mathcal{O}\pqty{\frac{1}{\beta_W}}\ .
\end{equation}
These relations can help compare studies with different actions, but they are not meant to be accurate. A comparison of the two couplings at some values of $\beta$ can be found in Table \ref{tab:wilson villain coupling comparison}. In particular, typical values used for simulations of the Wilson action are $\beta_W \in [1.7,2.4]$ which would correspond to $\beta_V \in [1.1, 1.8]$. This will inform our choice of parameters for numerical simulations, as explained in Section \ref{sec:numerical}.

With the Wilson action, monopoles can be defined according to the DeGrand-Toussaint prescription \cite{DeGrandToussaint}. With this prescription, one splits each plaquette variable into a $[0,2\pi)$-valued piece and an integer excess,
\begin{equation}
    d\varphi_p = (d\varphi_p \,\,\mathrm{mod}\,\, 2\pi) +
    2\pi m_p \ .
\end{equation}
The monopole number in a cube $c$ is then obtained by computing the flux of $m_p$ through the cube $c$, i.e. just $dm_c$. In other words, we can effectively use $m_p$ as the integer field in the Villain formulation.\footnote{By the arguments outlined in this and in the previous section, we then expect that imposing the constraint $dm=0$ in the Wilson action would lead to a massless theory much like it does for the Villain model.}

If we take the gauge field to lie in $[0, 2\pi)$, then we see that the monopole number can only take finitely many values, i.e. on the cubic lattice
\begin{equation}
    dm_c \in \{-2,-1,0,1,2\} \ .
\end{equation}
This usually does not cause any issues: since the monopole energy grows like $q^2$ where $q$ is the charge of the monopole, it is favourable for, say a charge $2$ monopole to decay to two charge $1$ monopoles (this process also preserves the total monopole number). Therefore monopoles of higher charge are usually suppressed. For $\mu \neq 0$, on the other hand, it is not clear that this argument still holds. The possibility that this restriction may affect the physics was raised in the previous paper on monopole enhancement \cite{athenodorou2026differentkindcontinuumlimit}. In later sections we show that restricting the monopole charges does not lead to qualitative differences unless the ratio $\mu /\beta$ is too large, in which case fluctuations on top of the background are suppressed.

\subsection{Predictions for \texorpdfstring{$\mu=0$}{the standard theory}}\label{sec:standard theory predictions}

In this section we review some of the theoretical predictions for the $\mu=0$ theory. We only give the final results, since their derivation will essentially be repeated during the derivation for the $\mu \neq 0$ case. The main references are the original derivation due to Polyakov \cite{Polyakov1977} and the rigorous work for the lattice theory due to G\"opfert and Mack \cite{GopfMack}. However we point the reader to several other useful references \cite{Banks:1977cc, Ambjorn:1982ts, Muller1981, SemiAbelian, MarianiThesis}.

The standard $\U(1)$ lattice gauge theory ($\mu=0)$ is confining for all values of the gauge coupling $\beta$ and predictions for the mass and string tension have been obtained via an effective theory. Some of these results, such as confinement, have been shown rigorously \cite{GopfMack}. In particular, the mass gap $m$ scales as (restoring the lattice spacing $a$) 
\begin{equation}
    \label{eq:mass gap}
    a^2 m^2 = 8\pi^2 \beta \exp{\pqty{-2\pi^2 G(0)\beta}} \ ,
\end{equation} 
where $G(0)$ is the value of the lattice Green's function $G$ at zero separation, i.e. for a cubic lattice \cite{Watson} 
\begin{align}
    \label{eq:G(0) cubic lattice}
    G(0) &= \frac{1}{2\pi^3} \int_{0}^{\pi} \int_{0}^{\pi}\int_{0}^{\pi}\frac{du \, dv \, dw}{3- \cos(u) - \cos(v) - \cos(w) } = \nonumber\\ &=\frac{\sqrt{6}}{192\pi^3} \Gamma{\pqty{\frac{1}{24}}} \Gamma{\pqty{\frac{5}{24}}} \Gamma{\pqty{\frac{7}{24}}} \Gamma{\pqty{\frac{11}{24}}} \approx 0.2527\ldots
\end{align}
On the other hand, the string tension $\sigma$ scales as (again, restoring the lattice spacing $a$)
\begin{equation}
    \label{eq:string tension analytical}
    a^2 \sigma = \frac{c}{4\pi^2 }\, \frac{a m}{\beta}  \ ,
\end{equation}
in terms of a dimensionless constant $c$. A semiclassical calculation leads to $c=8$. These predictions have found numerical confirmation~\cite{TepAthen, Loan2003, Loan:2003wy}. The constant $c \approx 8.29$ was estimated in \cite{TepAthen} for the Wilson action and is not far from the semiclassical value. As the continuum ($a\to 0$, $\beta \to \infty$) is approached, we then see that the dimensionless ratios involving mass and string tension behave as
\begin{equation}
    \label{eq:ratio scaling}
    \frac{\sigma}{m^2} \to \infty  \ , \quad \quad \quad \frac{(a^2 \sigma) \beta}{(am)} \to \frac{c}{4\pi^2} \ .
\end{equation}
Equation~\eqref{eq:ratio scaling} shows that the theory has several inequivalent length scales, even in the continuum. As such, in theory different continuum limits may be obtained depending on what quantity is taken as the standard of length. In a continuum limit where the mass $m$ is held fixed in physical units, in the large $\beta$ regime $a \to 0$ implies, via eq. \eqref{eq:mass gap} $\beta \to \infty$, and in particular $\beta(a) \sim -\log(a)$. The string tension becomes infinite, and the theory is equivalent to a free scalar (the \textquote{photoball}) of mass $m$ \cite{GopfMack}. This theory, while perhaps unusual, is not pathological. The prediction of a free scalar in the continuum limit can be checked numerically by computing a few low-lying glueball masses \cite{TepAthen} or at finite temperature by comparing the pressure to that of a free massive boson \cite{Caselle:2025_Pressure}.

If instead one holds fixed the string tension $\sigma$, then the mass (in physical units) goes to zero. This has important implications for the confining string of the theory. In particular there will be a regime where $m^2 \ll \sigma$, and therefore one cannot ignore interactions between the \textquote{photoball} and the transverse modes of the string. This leads to a non-universal form of the confining string potential \cite{Aharony:2024ctf} which has been checked numerically \cite{Caselle:2025_String}. Note that there is some doubt whether the continuum limit exists in a mathematical sense in this case. In particular the existence of the continuum limit was shown for the cases where either the mass \cite{GopfMack} or bare coupling \cite{GrossConvergence} are held fixed, but not for the string tension. 

Finally, one could hold fixed the dimensionful bare coupling $a\beta(a) \equiv 1/g^2$. The continuum theory is free electrodynamics \cite{GrossConvergence}, which is always in a massless phase.

\section{Analytical predictions}

In this section, we derive several analytical predictions for the monopole-enhanced theory. We first obtain the monopole gas representation for this model, and show that its ground state is given by an antiferromagnetically ordered Néel state. We then expand around the ground states and perform a Hubbard-Stratonovich transformation to obtain the theory of a real scalar field. After integrating out the short-distance fluctuations, we obtain an effective theory valid in the large $\beta$ regime. We then discuss some connections with other models, in particular as they relate to a prediction for a phase transition in the small or moderate $\beta$ range.

\subsection{Monopole gas representation}\label{sec:monopole gas}

In the presence of a monopole chemical potential it is no longer possible to perform Poisson summation and therefore it does not appear to be possible to obtain a dual theory. However, one can still obtain a monopole gas representation. While this derivation is entirely standard, we reproduce it here in order to convince the reader that no novelty arises out of the introduction of the chemical potential term.

The monopole gas representation arises by integrating out the gauge field $\varphi$. This can be done directly because the corresponding integral is Gaussian, and one is left with a theory of monopoles interacting via a long-range potential. The starting point is the Hodge decomposition (see Appendix \ref{sec:lattice differential forms}) for the integer field\footnote{Note that implicitly in infinite volume we're taking boundary conditions such that $m \to 0$ at infinity. Then one is not allowed to shift $m$ by a constant, otherwise the following equation would be inconsistent},
\begin{equation}
    \label{eq:hodge decomp main text}
    m = d \square^{-1} \delta m + \delta \square^{-1} d m \ ,
\end{equation}
where $\square=d\delta + \delta d$ is the lattice Laplacian. Here $d$ is the exterior derivative operator on lattice differential forms, while $\delta$ is the co-derivative. The action can then be rewritten as
\begin{align}
    \norm{ d\varphi_p+2\pi m_p }^2 &= \norm{ d (\varphi+2\pi \square^{-1} \delta m) + 2\pi \delta \square^{-1} d m }^2 = \nonumber\\
    &=\norm{ d (\varphi+2\pi \square^{-1} \delta m)}^2 + \norm{2\pi \delta \square^{-1} d m }^2 \ . \nonumber
\end{align}
In the above formula there is no cross-term, because it would take the form
\begin{equation}
    (da, \delta b)=(d^2a, b)=0 \ .
\end{equation}
Now we can absorb the shift into $\varphi$ by redefining,\footnote{If we had chosen $\varphi_l \in [0,2\pi)$, then this redefinition makes $\varphi_l$ non-compact and therefore the final result is the same. The argument is slightly more involved: one would show that it is possible to sum separately over $dm$ and $\delta m$ due to the Hodge decomposition. }
\begin{equation}
    A = \varphi + 2\pi \square^{-1} \delta m \ .
\end{equation}
Then the partition function factorizes into that of real electrodynamics and that of the monopole gas,
\begin{equation}
    Z=Z_\R Z_M \ ,
\end{equation}
where
\begin{equation}
    Z_\R = \pqty{\prod_l\int_\R dA_l} \exp\bqty{-\frac{\beta}{2} \sum_p (dA_p)^2} \ ,
\end{equation}
and on the other hand, 
\begin{equation}
    Z_M =  \pqty{\prod_p\sum_{m_p \in \Z}} \exp\bqty{-2\pi^2\beta \sum_p (\delta \square^{-1} d m)_p^2+\mu\sum_c \abs{dm_c}} \ .
\end{equation}
This last expression can be simplified, since $dm$ is a $3$-form and therefore automatically closed. This means that $\square^{-1} dm$ is also a $3$-form, and therefore automatically closed. On an arbitrary $3$-form $\alpha$, one then has
\begin{equation}
    \square \alpha = (d\delta+\delta d) \alpha = (d\delta) \alpha \ ,
\end{equation}
so that $d\delta \square^{-1} \alpha =\alpha$. Therefore, 
\begin{align*}
    \norm{\delta \square^{-1} dm}^2 &= (\delta \square^{-1}dm, \delta \square^{-1}dm)\\
    &= ( \square^{-1}dm, d \delta \square^{-1}dm) = ( \square^{-1}dm, dm)
\end{align*}
So that we find
\begin{equation}
    Z_M =  \pqty{\prod_p\sum_{m_p \in \Z}} \exp\bqty{-2\pi^2\beta  (dm,  \square^{-1} d m)+\mu\sum_c \abs{dm_c}} \ .
\end{equation}
Now since the action depends only on $dm$, we can replace $n = dm$. In infinite volume this is always possible, while in finite volume this is true as long as the constraint $\sum n = \sum dm =0$ is satisfied. Therefore we can replace integration over $m$ with integration over $dm$. For simplicity we also go to the dual lattice so that we work with sites rather than cubes and we find
\begin{equation}
    Z_M =  \pqty{\prod_x\sum_{n_x \in \Z} \delta_{\sum n}} \exp\bqty{-2\pi^2\beta  (n,  \square^{-1} n)+\mu\sum_x \abs{n_x}} \ .
\end{equation} 
which completes the derivation. 

We also briefly derive the expression for a Wilson loop expectation value in the monopole gas representation. This will be used later to obtain the string tension. In the original gauge theory the Wilson loop observable takes the form
\begin{equation}
    \exp(i(\gamma, \varphi)) \ ,
\end{equation}
where $\gamma_l = \pm 1$ (depending on relative orientation) on the links in the loop, and zero otherwise. It is easy to see that $\delta \gamma = 0$ (for a simple loop, each point has one ingoing and one outgoing term), and that moreover $\gamma =\delta \omega$, where $\omega=\pm 1$ on the plaquettes belonging to a surface bounded by the Wilson loop, and zero otherwise. We choose the minimal surface. Then integration by parts leads to the following expression for the Wilson loop observable,
\begin{equation}
    \exp(i(\omega, d\varphi)) \ .
\end{equation}
With the introduction of the variable $A$, one then has 
\begin{equation}
    \exp(i (\omega, d\varphi))=\exp(i (\omega,dA)) \exp(-2\pi i (\omega,d \square^{-1} \delta m)) \ .
\end{equation}
Therefore the observable factorizes into the product of a non-compact part and a monopole part, much like the partition function. As in the standard theory, the non-compact part of the observable can be evaluated exactly and gives a subleading logarithmic contribution (i.e. not area law). For the other piece, we go back to the Hodge decomposition for $m$ so that we can express the observable in terms of $dm$. Since both $m$ and $\omega$ are integer-valued, the $m$ component in the expression is one and, after using integration by parts, we obtain the following expression for the monopole part of the Wilson loop observable,
\begin{equation}
    \label{eq:wilson loop observable monopole representation}
    \exp( 2\pi i (\eta, n)) \ ,\qquad \star\eta = \square^{-1} d \omega \ .
\end{equation}
where $n$ and $\eta$ now live in the dual lattice, and the $\star$ operator relates forms on the primal and dual lattices. On the dual lattice, one can also write $\eta = \square^{-1} \delta \star \omega$, where $\star \omega=\pm1$ on links dual to plaquettes in the minimal surface bounded by the loop, and zero otherwise.

\subsection{Monopole ground state and excitations}\label{sec:monopole ground state}

In this section, we study the monopole gas in the presence of the chemical potential term. We first obtain the ground state of the system, which dominates for sufficiently large $\beta$ and then study the excitations on top of this. For convenience, we write
\begin{equation}
    \mu \equiv \kappa \beta \ ,
\end{equation}
since we care mostly about the positive $\mu$ regime. Note that we do not require $\kappa$ to be constant; eventually we'll be interested in sending $\beta \to \infty$ and we allow arbitrary trajectories $\kappa=\kappa(\beta)$. With this replacement, the monopole gas partition function can be written as 
\begin{equation}
    Z_M =  \pqty{\prod_x\sum_{n_x \in \Z} \delta\pqty{\sum n}} \exp\bqty{-\beta E_M(n)} \ ,
\end{equation}
where
\begin{equation}
    E_M(n)=2\pi^2  (n,  \square^{-1} n)-\kappa \sum_x \abs{n_x} \ .
\end{equation}
For large $\beta$, the system is dominated by the ground state. This can be found as follows. Note that on the cubic lattice, the Laplacian is diagonal in momentum space and given by
\begin{equation}
    \label{eq:lattice Laplacian momentum space}
    \square(p)=4 \sum_{\mu=1}^3 \sin^2(p_\mu/2) \ .
\end{equation}
It has a zero mode corresponding to constant functions, but these are disallowed by the neutrality constraint. Then we see that in fact
\begin{equation}
    \square(p) \leq 12 \ ,
\end{equation}
with equality if and only if $p_\mu=\pi$ for $\mu =1,2,3$ (up to periodicity). This then implies
\begin{equation}
    \label{eq:lattice Green's function bound}
    \square^{-1} \geq \frac{1}{12} \ .
\end{equation}
In other words, as a matrix, its smallest eigenvalue is $1/12$. Then we can use the well-known matrix inner product inequality, $x^T A x \geq \lambda_{\mathrm{min}} x^Tx$ to obtain a lower bound
\begin{equation}
    \label{eq:monopole ground state inequality}
    E_M \geq \sum_x \bqty{\frac{\pi^2}{6}  n_x^2 -\kappa \abs{n_x}}  \ .  
\end{equation}
The quantity in brackets is minimized when
\begin{equation}
    \abs{n_x} = \frac{3\kappa}{\pi^2} \ .
\end{equation}
This absolute minimum is achievable only if $\kappa$ is such that the relevant value is an integer; otherwise it is achieved at the closest integer to the above value. Therefore for clarity of notation set
\begin{equation}
    \frac{3\kappa}{\pi^2} \equiv q \in \R_{+} \ .
\end{equation}
Then the ground state is located at
\begin{equation}
    \abs{n_x}=[q] \ ,
\end{equation}
where $[\cdot]$ is the closest integer function. From now on, we will use $q$ as a real parameter due to its direct connection with the ground state. Then if $n_x \neq 0$, the inequality being valid as an equality only for the state of momentum $p_\mu=\pi$, we see that the ground state is in fact given by
\begin{equation}
    n_x = \begin{cases}
        (-1)^x[q] & q > 1/2\\
        0         & 0 < q < 1/2
    \end{cases}
\end{equation}
This implies, in particular, that for sufficiently large positive chemical potential, the ground state of the monopoles is a dense monopole crystal arranged antiferromagnetically. The monopole crystal spontaneously breaks both the $\Z_2$ charge conjugation symmetry as well as translations by one lattice spacing. For small values of the chemical potential, we have the usual dilute phase. The ordering of the monopoles can be detected by the staggered magnetization
\begin{equation}
    \label{eq:staggered magnetization}
    M_s = \frac{1}{V} \sum_{x} (-1)^x n_x \ .
\end{equation}
For small $\beta$, we expect the system to be disordered with $M_s=0$. For large $\beta$, instead, the system should order and we expect instead $M_s=\pm [q]$. In Section \ref{sec:long-range ising} we explain why we expect this transition to be continuous and  belong to the short-range Ising universality class. 

Instead, at half-integer $q \in \Z_{\geq 0}+\frac12$ there is an ambiguity in the ground states. It is easy to check that at those points the energies of the two competing ground states $\lfloor q\rfloor$ and $\lceil q\rceil$ is the same,
\begin{equation}
    E_M((-1)^x \lfloor q\rfloor) = E_M((-1)^x\lceil q\rceil) \ .
\end{equation}
At these points, we therefore expect phase coexistence which is a sign of a first-order transition. Of course this holds for large enough $\beta$; for smaller $\beta$, instead, we expect the system to remain in a dilute phase. A summary of the phase diagram, incorporating also information from the next sections, can be found in Section \ref{sec:analytic summary}.

We then look at the excitations on top of the dense monopole crystal background. In the dilute case, the typical excitation is a monopole-antimonopole pair over the neutral background (so as to preserve overall neutrality). In this case, instead, we have two types of excitations:
\begin{enumerate}
    \item A pair of background charges flip.
    \item A monopole-antimonopole pair is created over the dense background.
\end{enumerate}
Of course a pair of spin flips can be considered as a \textquote{large} excitation of the latter type, but it is conceptually important to distinguish the two. To better compare with the standard theory ($\mu=0$), we first compute the energy cost of a monopole-antimonopole in the standard theory. Suppose we put a monopole of charge $k >0$ at position $x$ and an antimonopole of charge $-k$ at position $y$. In other words, we consider the configuration
\begin{equation}
    \label{eq:monopole excitation}
    \Delta n_z = k(\delta_{xz}-\delta_{yz}) \ . 
\end{equation}
Then the energy above the vacuum $n_z\equiv0$ is given by ($r=\abs{x-y})$
\begin{equation}
    \Delta E_{k,-k} = 4\pi^2 k^2(G(0)- G_{xy}) \approx 4\pi^2 k^2\pqty{G(0)-\frac{1}{4\pi r}} \ .
\end{equation}
Here we used the fact that in three dimensions the lattice Green's function $G\equiv \square^{-1}$ at zero separation $G(0)$ is finite, and for non-zero separation it approximates the Coulomb potential. This shows in particular that the excitations are mostly $k=1$ and their energy slowly increases with distance. 

Now for $\mu > 0$ suppose we are in the sector dominated by charge $[q]$ monopoles. Consider an excitation of the same type as above, i.e. given by eq.\eqref{eq:monopole excitation}. In this case the vacuum is $(n_0)_x =s (-1)^x [q]$ with $s=\pm 1$ the global sign. Using the fact that $n_0$ is an eigenvector of $G$, we obtain 
\begin{multline}
    \Delta E_{k,-k} = 4\pi^2 (\delta n, G n_0) + 2\pi^2 (\delta n , G \delta n)\\ -\kappa \sum_x \pqty{\abs{n_0+\delta n}-\abs{n_0}} \ .
\end{multline}
Now we distinguish between \textquote{small} excitations, with $k < [q]$ and \textquote{large} excitations with $k > [q]$. In the small case, we find
\begin{multline}
    \Delta E_{k,-k} = \frac{\pi^2 sk}{3} \pqty{[q] - q} \pqty{(-1)^x-(-1)^y} \\+ 4\pi^2 k^2 (G(0)-G(r)) \ .
\end{multline}
The most important piece of information from this last equation is that, exactly at the integer point, the excitations look exactly like the excitations in the standard theory. Therefore, one would expect that they are similarly dilute and weakly-coupled. Away from the integer point, there are corrections proportional to $q-[q]$. But these corrections should not affect the qualitative structure of the theory, since we are in the same phase (although, of course, they have to exist because of the coexistence point where the ground state is a soup of different charges).

For $x$ and $y$ on the same sublattice, the excitations have the same energy as in the standard theory for any $\kappa$. Instead for opposite sublattices, the correction is proportional to $[q]-q$ and therefore which excitations are preferred depend (unsurprisingly) on which integer charge we're closest to. For example, if $[q]-q<0$, then since we are assuming $(-1)^x (-1)^y = -1$, the excitation which is favored has $(-1)^x s=1$, i.e. if the monopole background at site $x$ is $+[q]$ it adds $+k$ positively, and oppositely on oppositely charged sites.

On the other hand, \textit{large} excitations have $k \geq [q]$. In this case, we find
\begin{multline}
    \Delta E_{k,-k} = \frac{\pi^2}{3} [q]\pqty{ k-q} \pqty{s(-1)^x-s(-1)^y}\\ -2\kappa (k-[q]) + 4\pi^2 k^2 (G(0)-G(r))  \ .
\end{multline}
Here the excitations are different from usual even in the same sublattice case, since there seems to be a competition between the $k^2$ term suppressing high monopoles and the $-k$ term wanting to increase them. Later we will see that these excitations are subleading.

\subsection{Effective theory description}\label{sec:effective theory}

In this subsection we derive an effective theory description for the $\mu >0$ theory at large $\beta$, similarly to Polyakov's derivation for the standard theory. The standard argument is based on the observation that the monopole gas is dilute and weakly-coupled. Based on what we have learned in the previous sections, for sufficiently large $\mu$, the monopoles are arranged in an antiferromagnetic Néel state. But the excitations on top of this dense background are very similar to those of the ordinary theory. This gives a natural entry point for the derivation of the effective theory: in particular, one introduces a scalar field for the \textquote{spin wave} excitations on top of the dense monopole background rather than on top of the neutral background. Another novel ingredient compared to the standard derivation is the need to integrate out short-range fluctuations due to the antiferromagnetic order. The final effective theory is a sine-Gordon model which allows us to obtain predictions for both the mass and string tension.

The starting point is the decomposition of the monopole state into dense background plus excitations, i.e. 
\begin{equation}
    \label{eq:monopole decomposition}
    n_x = s_x [q] + l_x \ ,
\end{equation}
where $s_x = s (-1)^x $ corresponds to staggering and $l_x \in \Z$ to the excitations; these again satisfy a neutrality condition, $\sum_x l_x=0$. Substituting into the energy, we find
\begin{equation}
    E_M(l) = 2\pi^2(l, \square^{-1} l) + \sum_x V(s_x l_x)  \ ,
\end{equation}
where
\begin{equation}
    V(s_x l_x) = \frac{\pi^2}{3}q [q] +  \frac{\pi^2}{3} ([q]s_xl_x- q\abs{[q] + s_x l_x}) \ .
\end{equation}
In the previous equation we have added a constant to $V$ in such a way that
\begin{equation}
    V(0)=0 \ .
\end{equation}
This will lead to some simplifications later on. 

Now we introduce a scalar field via a Hubbard-Stratonovich transformation, which will also turn the long-range interaction between monopoles into a short-ranged interaction. In order to make the resulting sum over monopoles convergent, we adopt a Pauli-Villars regulator as in the standard theory \cite{GopfMack} and write
\begin{equation}
    \label{eq:PV split}
    \square^{-1} = \square^{-1}_{\mathrm{PV}} + \square^{-1}_M \ ,
\end{equation}
where $G_M\equiv \square^{-1}_M = (\square +M^2)^{-1}$ is a massive Green's function and $\square^{-1}_{\mathrm{PV}}$ is given by subtraction. Now a standard argument would be that due to the large mass $M$ which we'll eventually send to infinity, we can make the replacement\footnote{\label{foot:gopfmack clarification}Note that this argument, while standard, is not quite correct as stated. In their rigorous work on the standard $\U(1)$ theory, G\"opfert and Mack \cite{GopfMack} choose $M$ so that there is a hierarchy of scales,  
\begin{equation}
    \label{eq:PV hierarchy}
    m \ll M \ll a^{-1} \ . 
\end{equation}
where $m$ is the mass gap eq.\eqref{eq:mass gap}. This then means that the interaction range of the massive propagator, $1/M$, is much smaller than the physical correlation length $1/m$, but much larger than the lattice spacing, $1/M \gg a$. Thus one would need to do more work to show that in fact one can replace $G_M(x-y)$ with $G_M(0)$. Here we limit ourselves to noting that, as we have seen in Section \ref{sec:monopole ground state}, the structure of the excitations on top of the dense background is the same as in the standard theory. Since $G_M$ couples to these excitations, we expect that this approximation can be made rigorous also in our model, but we emphasize that our arguments do not provide a rigorous proof of this fact.

Finally, we comment on the validity of the replacement $G_M(0) \to G(0)$. From its definition one has an explicit formula from which it is not hard to see that $G_M(0) \approx G(0) + \mathcal{O}(M)$. Then note that $M$ can be chosen to further satisfy $\beta M \to 0$ as $\beta \to \infty$ (for example $M=\sqrt{\beta} m$ satisfies this property as well as the hierarchy eq.\eqref{eq:PV hierarchy}). Since $G_M$ appears in all formulas multiplied by $\beta$, the correction $\mathcal{O}(\beta M)\to 0$ as $\beta \to \infty$.}
\begin{equation}
    (l, G_M l) \approx \sum_{x} G(0) l_x^2 \ .
\end{equation}
We then perform a Hubbard-Stratonovich transformation, recovering a real scalar field, but only for the PV part, i.e. we write
\begin{multline}
    \exp\bqty{-2\pi^2\beta  (l,  \square_{\mathrm{PV}}^{-1} l)} = \pqty{\prod_x \int_\R d\phi_x}  \times \\
    \times \exp{\pqty{-\frac{1}{8\pi^2\beta} (\phi,\square_{\mathrm{PV}}\phi) -i(\phi,l) }} \ .
\end{multline}
With the approximation described above, we get\footnote{In this expression, we have ignored the monopole neutrality constraint enforced by the delta function. This is because it can be written in terms of integration over a global compact variable $\theta$,
\begin{equation}
    \delta\pqty{\sum n} = \frac{1}{2\pi}\int_{-\pi}^{\pi} d\theta \, e^{i \theta \sum n}  \ .
\end{equation}
Then $\theta$ can be absorbed into a shift of $\phi_x$, which corresponds to the zero mode of the kinetic term. Therefore monopole neutrality (which ensured the invertibility of the Laplacian) is related to the zero mode of $\phi$, which must be removed to ensure the invertibility of the kinetic term but which we can also ignore for our purposes. }
\begin{multline}
    Z_M = \pqty{\prod_x \int_\R d\phi_x} \exp{\pqty{-\frac{1}{8\pi^2\beta} (\phi,\square_{\mathrm{PV}}\phi)}}  \times \\ \times  \prod_x \pqty{\sum_{l \in \Z} \exp{\pqty{-i\phi_x l -2\pi^2 \beta G(0) l^2 -\beta V(s_x l) }} } \ .
\end{multline}
Now we can push the staggering onto $\phi_x$ by performing a change of variables $s_x l \to l$, which leads to 
\begin{multline}
    Z_M = \pqty{\prod_x \int_\R d\phi_x} \exp{\pqty{-\frac{1}{8\pi^2\beta} (\phi,\square_{\mathrm{PV}}\phi)}}\times \\
    \times \prod_x \pqty{\sum_{l \in \Z} \exp{\pqty{-i\phi_x (-1)^x l -2\pi^2 \beta G(0) l^2 -\beta V(l) }} } \ .
\end{multline}
Now for clarity we replace the label $l$ with $n$. From now on, call
\begin{equation}
    W = \prod_x \pqty{\sum_{n \in \Z} \exp{\pqty{-i\phi_x (-1)^x n -2\pi^2 \beta G(0) n^2 -\beta V(n) }} } \ . 
\end{equation}
the term which will give rise to a potential.

Now, as is standard in theories of antiferromagnets \cite{Affleck, ReadSachdev,Staggered} progress can be made by integrating out the short distance fluctuations. This will remove the staggering factors $(-1)^x$ which are cumbersome near the continuum. The standard way of doing so involves passing to sum and difference variables,
\begin{align}
    \label{eq:sum and difference variables}
    \chi_x &= \frac{\phi_x + \phi_{x+\hat{1}}}{2} \ ,\\
    \xi_x &= \frac{\phi_x - \phi_{x+\hat{1}}}{2} \ .
\end{align}
In this expression, we have singled out an arbitrary direction, but in the final effective theory this explicit breaking will be irrelevant. Note that it is key that the sum and difference variables be introduced for the scalar field; the map eq.\eqref{eq:sum and difference variables} has constant Jacobian, while for the integer-valued monopole variables the corresponding transformation would have been more complicated. 

The sum and difference fields $\chi, \xi$ live only on half the lattice sites. By convention, we choose them to live on the even sites of the cubic lattice, i.e. those with $(-1)^{x+y+z}=1$. The corresponding lattice (i.e. the sublattice of even sites of the cubic lattice) forms a face-centered cubic (FCC) lattice, whose corresponding  tessellation of three-dimensional space is known as the tetrahedral-octahedral honeycomb. A standard result relates a standard kinetic term for $\phi$ to those of $\chi, \xi$, i.e. \cite{ReadSachdev, Staggered}
\begin{align}
    \label{eq:standard kinetic term sum and difference}
    \sum_{\expval{xy}} \frac12 (\phi_x-\phi_y)^2 
    &\approx \sum_{x\,\mathrm{even}} \bqty{ 12 \xi_x^2 + \frac12 (\nabla \chi_x)^2} \,
\end{align}
where $\nabla\chi$ is a standard kinetic term for the $\chi$ field in the FCC lattice. A derivation of this result is also given in Section \ref{sec:sum and difference PV}. The key point is that $\chi$ (the sum field) couples to the low momentum modes, while $\xi$ (the difference field) couples to the corner of the Brillouin zone and is massive (the factor of $12$ is the same factor that we found in eq.\eqref{eq:lattice Green's function bound}). In particular, since $\xi$ has a mass of the order of the inverse lattice spacing, we can neglect its fluctuations. Thus we can neglect both its kinetic term and other Lorentz symmetry violating terms which would have otherwise appeared due to the arbitrary choice of direction in eq.\eqref{eq:sum and difference variables}. As we will see in more detail in what follows and in particular in Section \ref{sec:sum and difference PV}, it would be more appropriate to consider the full Pauli-Villars regulated operator rather than the standard kinetic term for $\phi$, which leads to a small but significant difference in the final answer.

In the sum which will produce the potential term, we now sum only over even $x$, and replace sum and difference variables, so that we get
\begin{multline}
    W=\prod_{x\, \mathrm{even}}\Bigg( \sum_{n, m \in \Z} \exp{\pqty{-i (\chi_x+\xi_x) n + i (\chi_x-\xi_x) m}}\times \\ \times  \exp{\pqty{ -2\pi^2 \beta G(0) (n^2+m^2) -\beta (V(n)+V(m) }}\Bigg)  \ . 
\end{multline}
Now we can simply integrate out $\xi$. With the approximation of eq.\eqref{eq:standard kinetic term sum and difference}, this can be done easily and one finds, term by term,
\begin{equation}
    \int d\xi_x \, e^{-\frac{3}{\pi^2 \beta} \xi_x^2} \exp{\pqty{-i \xi_x (n +m )}} = \exp{\pqty{-\frac{\pi^2 \beta}{12}(n+m)^2 }} \ .
\end{equation}
If instead one considers the full PV-regulated propagator, we will see that this term is in fact equal to one up to exponentially small corrections. The point is that by replacing the PV kinetic term with a standard kinetic term we are \textquote{double counting} the contribution due to $G(0)$ which we had previously accounted for separately. To avoid overcomplicating the discussion, we present the calculation with the full Pauli-Villars kinetic term in Section \ref{sec:sum and difference PV} rather than here. In the terms of the final formulas, this will make a small but significant difference in the exponent for the mass gap, and one realizes that it must be there by comparing with the result for the $\mu=0$ theory. The final result is that 
\begin{equation}
    W=\prod_{x\, \mathrm{even}} \pqty{\sum_{n, m \in \Z} \exp{\pqty{-i \chi_x (n-m)}} y_n y_m} \ , 
\end{equation}
where
\begin{equation}
    y_n = \exp(-2\pi^2 \beta G(0)  n^2-\beta V(n)) \ .
\end{equation}
Now we write $D=n-m$ and collect terms by the value of $D$, i.e. we write
\begin{equation}
    \label{eq:K(D) definition}
    K(D) = \sum_{\substack{n,m \in \Z\\ n-m=D}} y_n y_m \ .
\end{equation}
Because the sum in $D$ is symmetric under the exchange of $n$ and $m$, we have that
\begin{equation}
    K(D)=K(-D) \ .
\end{equation}
This then leads to a manifestly real expression for the sum, 
\begin{equation}
    W=\prod_{x\, \mathrm{even}} \pqty{K(0)+2\sum_{D =1}^{\infty} K(D) \cos{\pqty{\chi_x D}}} \ .
\end{equation}
Now comes a key point of the argument. We can only interpret $W$ as giving rise to a potential for the scalar field $\chi$ if it is strictly positive, $W > 0$, so that the potential for $\chi$ is $U(\chi)=-\log(W)$. \footnote{Note that this is precisely the failure point of a spin wave expansion around the trivial monopole vacuum for the $\mu > 0$ theory: one obtains a mathematically well-defined expression for the partition function, but with a complex-valued potential. In other words, with the onset of a non-trivial monopole background the analogue of $W$ becomes negative.} It turns out that indeed for each separate $x$, the series appearing in $W$ is positive. Moreover, one can find an approximate expression for $U(\chi)$ up to exponentially small corrections in $\beta$. Since the proof is somewhat long, it can be found in Section \ref{sec:positivity proof} together with the approximation for $U$. The final result is that we find an effective theory for the scalar field $\chi$ and partition function
\begin{equation}
    Z_M = \pqty{\prod_{x\, \mathrm{even}} \int_\R d\chi_x} \exp{\pqty{-S[\chi]}} \  ,
\end{equation}
where, canonically normalizing the field, the action is given by
\begin{equation}
    S[\chi] = 2\sum_{x\, \mathrm{even}} \bqty{\frac{1}{2} (\nabla \chi_x)^2 +z(\beta, q)(1-\cos(2\pi\sqrt{\beta}\chi_x))  }\ .
\end{equation}
Here we gathered a factor of $2$ outside, since in lattice units the unit cell of the FCC lattice has twice the volume as that of the original cubic lattice. Therefore in the continuum limit, $2\sum_{x \in \mathrm{even}} \to \int d^3 x$. Note also that $(\nabla \chi_x)^2$ is to leading order just the standard kinetic term on an FCC lattice, which converges to a standard kinetic term in the continuum limit. The prefactor in the potential is given by
\begin{equation}
    z(\beta, q) = 2\exp(-2\pi^2 G(0)\beta)\begin{cases}
         \cosh\pqty{\frac{\pi^2}{3}\beta ([q]-q)} & q >  1/2\\
         \exp\mu & q < 1/2
    \end{cases}
\end{equation}
We are mostly interested in the case $q > 1/2$, but the other case $q < 1/2$ is useful as a cross-check which indeed matches the predictions discussed in Section \ref{sec:standard theory predictions}. For the $q > 1/2$ theory, expanding the cosine to leading order allows us to extract a mass
\begin{align}
    \label{eq:mass gap pediction}
    a^2m^2 &= 4\pi^2 \beta z(\beta, q) =\\ 
    &=8 \pi^2 \beta e^{-2\pi^2G(0)\beta} \cosh\pqty{\frac{\pi^2}{3}\beta ([q]-q)} \ .
\end{align}
Note that for $\mu=0$ we find the same formula for the mass as in the standard theory, eq.\eqref{eq:mass gap}. Moreover, at the integer points $q=[q]$ the mass gap is again the same as in the standard theory. This result is also qualitatively consistent with the observation in \cite{athenodorou2026differentkindcontinuumlimit} that the mass gap increases for $\mu > 0$.

The remaining step is the calculation of the string tension, so we can check the behavior of the mass to string tension ratio for $\mu > 0$. We have already derived the monopole gas representation of a Wilson loop observable in the original theory, given in eq.\eqref{eq:wilson loop observable monopole representation}. The first step is to check whether the dense monopole background gives rise to a shift in the string tension via eq.\eqref{eq:monopole decomposition}. By substituting the background explicitly in eq.\eqref{eq:wilson loop observable monopole representation}, we see that the dense background gives a contribution equal to 
\begin{equation}
    \exp(\pm 2\pi [q] i (\eta, (-1)^x)) \ .
\end{equation}
Recalling that we had $\eta = \square^{-1} \delta \star \omega $, and the fact that that the background is an eigenfunction of the lattice Laplacian, we find that
\begin{align}
    (\eta, (-1)^x) &= (\square^{-1}\delta  \star \omega, (-1)^x) =\\
    &=(\star \omega, d \square^{-1}(-1)^x)= \frac{1}{12} (\star \omega, d (-1)^x)
\end{align}
It is clear that this expression gives at most a perimeter law contribution. In particular, $d(-1)^x$ is a $1-$form living on lattice links equal to $\pm 2$ depending on the orientation of the links versus the staggering. When computing the inner product with $\star \omega$, any two neighboring links with $\omega \neq 0$ will share the same orientation but opposite values of $d(-1)^x$ due to the staggering. Hence all internal contributions can be paired up and vanish, leaving at most a contribution coming from the boundary terms, which is at most perimeter law. Thus the contribution due to the monopole background can be ignored. 

The contribution due to the monopole excitations can instead be written in terms of the $l$ variables, and it is given by
\begin{equation}
    \exp(2\pi i (\eta, l)) \ .
\end{equation}
Hence in the Hubbard-Stratonovich step it enters as a shift of
\begin{equation}
    \label{eq:phi shift due to wilson loop}
    \phi_x \to \phi_x -2\pi \eta_x \ .
\end{equation}
This then enters each sum over $n, m$ as
\begin{equation}
    \exp(2\pi i(\eta_x n -  m \eta_{x+\hat{1}})) \ .
\end{equation}
Now consider the error from replacing $\eta_{x+\hat{1}}$ with $\eta_x$ in the above formula. Note that
\begin{equation}
    \eta_x = \sum_y G(x-y) (\delta \star\omega)_{y}\  .
\end{equation}
Here $(\delta \star\omega)_y$ (morally a $\delta'$ in the continuum) equals $\pm 1$ on the sites on the two sides of the minimal area of the original Wilson loop (which lives in the dual lattice) and is zero otherwise. Then by translation invariance,
\begin{equation}
    \label{eq:eta difference}
    \eta_x-\eta_{x+\hat{1}} = \sum_y G(x-y) \bqty{(\delta \star\omega)_{y}-(\delta \star\omega)_{y+\hat{1}}} \ .
\end{equation}
Now we choose the Wilson loop to lie in a plane containing direction $\hat{1}$. The direction of the loop is arbitrary since, to leading order, the effective theory is isotropic, and therefore the resulting string tension is isotropic up to subleading corrections. With this choice, the difference in the square brackets in eq. \eqref{eq:eta difference} is non-zero only on two strips of width two sites in the $\hat{1}$ directions, i.e. it is supported only on the loop perimeter. Therefore the corresponding potential $\eta_x-\eta_{x+\hat{1}}$ gives at most a logarithmic contribution to the potential. Incidentally, it takes a very similar form as the contribution due to a non-compact field, which we saw at the end of section \ref{sec:monopole gas} and is again logarithmic. Therefore, it does not affect the string tension and can be neglected. The remaining term is then absorbed by a shift
\begin{equation}
    \chi_x \to \chi_x -2\pi \eta_x\  ,
\end{equation}
which is how a Wilson loop observable enters the dual sine-Gordon model in the standard theory. Then one is left to do a standard semiclassical calculation for the string tension in the sine-Gordon model, which can be found for example in \cite{MarianiThesis, Ambjorn:1982ts, Muller1981}. The final result is that \begin{equation}
    \label{eq:string tension prediction}
    a^2\sigma = \frac{2am}{\pi^2 \beta} \ ,
\end{equation}
which is the same equation as in the standard theory. Therefore also in this case,
\begin{equation}
    \frac{m^2}{\sigma}\to 0
\end{equation}
as $\beta \to \infty$ at fixed $\mu/\beta$, even though the dependence of $m$ on $\beta$ is somewhat different. As explained in detail for example in Chapter 2 of \cite{MarianiThesis}, the continuum limit with $m$ fixed is a massive real scalar.

\subsubsection{Proof of positivity and estimates}\label{sec:positivity proof}

In this section, we discuss the proof of positivity of the series $W$ which defines the scalar field potential in the effective theory. Due to its length, we had postponed this discussion from the main text of Section \ref{sec:effective theory}. 

The key intuition is that, having removed the monopole background as in eq.\eqref{eq:monopole decomposition}, the $l=0$ term is now dominant. Recalling the definition of the relevant quantities, let
\begin{equation}
    W(\chi_x) = K(0) +2\sum_{D =1}^{\infty} K(D) \cos(\chi_x D) \ .
\end{equation}
where $K(D)$ is defined as in eq.\eqref{eq:K(D) definition}. The main goal of this section is to show that in fact $W$ is strictly positive for all $\chi$ as long as $\beta$ is sufficiently large, i.e.
\begin{equation}
    W(\chi)>0 \ .
\end{equation}
The first step consists in noting that $\cos(\chi_z)\geq -1$ and therefore
\begin{equation}
    W(\chi_x) \geq K(0) -2\sum_{D =1}^{\infty} K(D) \ .
\end{equation}
Note that the cosines cannot all take value $-1$, therefore the inequality would be strict for the partial sums, but the limit procedure might in principle restore equality. Therefore, a sufficient condition for the positivity of $W$ is that 
\begin{equation}
    K(0) >2\sum_{D =1}^{\infty} K(D) \ .
\end{equation}
This essentially also shows that the series is dominated by the $l=0$ term, but we consider this more carefully at the end of the section. Now writing
\begin{equation}
    y_n = \exp{(-2\pi^2\beta G(0) n^2-\beta V(n))} \ ,
\end{equation}
we see by comparing with its definition eq.\eqref{eq:K(D) definition} that\footnote{If we had instead retained the term $(n+m)^2$ resulting from the $\xi$ integration due to the naive kinetic term, $K(D)$ would have been smaller than the right hand side and the argument goes through in the same way.}
\begin{equation}
    K(D) = \sum_{n,m \in \Z, n-m=D} y_n y_m \ .
\end{equation}
Since $y_0=1$ and all the other terms are positive, we see that 
\begin{equation}
    K(0) > 1 \ .
\end{equation}
Therefore it suffices to show that 
\begin{equation}
    2\sum_{D =1}^{\infty} K(D) < 1
\end{equation}
It is important to note that it suffices for our purposes that this is satisfied for sufficiently large $\beta$. 

\begin{figure}
    \centering

    \def\newpi{1.4}
    
    \begin{tikzpicture}[scale=0.9]
    \draw[->] (-3.5,0) -- (2.3,0) node[right] {$x$};
    \draw[->] (0,-3.0) -- (0,1.0) node[above] {$y$};
    
    \draw[blue, thick, domain=-3.3:-1, samples=100]
        plot (\x,{2*\newpi*\newpi/3*(1+\x)});
    \draw[blue, thick] (-1,0) -- (2,0);
    \node[blue, above right] at (0.3,0) {$V(x)$};

    \draw[red, thick, dashed, domain=-3.0:3.0, samples=150]
        plot (\x,{-\newpi^2/6*\x*\x});
    \node[red, below right] at (1.2,-0.8)
        {$y=-\frac{\pi^2}{6}x^2$};

    \draw (-1,0.1) -- (-1,-0.1)
        node[above] {$-q$};

\end{tikzpicture}
    \caption{Plot of $V(n)$ (\textit{smooth blue}) together with the optimal parabola lying entirely below it (\textit{dashed red}), in the case $q=1$. The figure is qualitatively the same for any integer $q$. In the non-integer case, the situation is more complicated.}
    \label{fig:V(n) and parabola}
\end{figure}
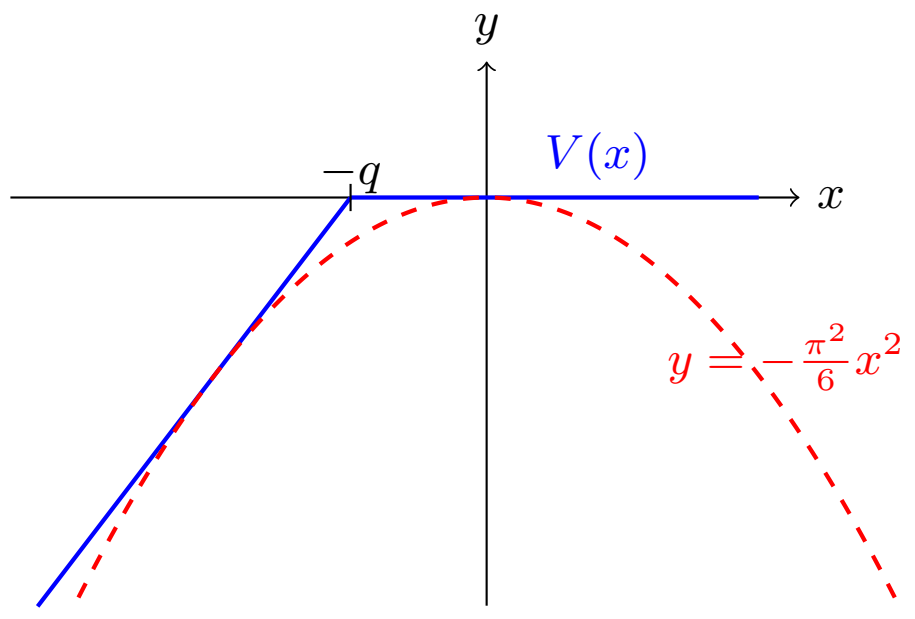

It is clear that now we must proceed by estimating the sum over $y_n$, which requires an upper bound on $y_n$. The difficulty lies mostly with $V(n)$ which gives a contribution increasing with $n$. But since this contribution is only linearly rising as opposed to the other term, which decreases quadratically, we can accommodate the increasing contribution into a quadratic bound by weakening its coefficient. Concretely, we have the inequality
\begin{equation}
    V(n) \geq -\frac{\pi^2}{6}n^2 \ .
\end{equation}
That such an inequality should exist is clear in the case where $q$ is an integer by plotting $V(n)$, since in this case it takes a particularly simple form,
\begin{equation}
    V(n)= \begin{cases}
       0  & n \geq -q\\
       \frac{2\pi^2 q}{3} (q + n)  & n \leq -q
    \end{cases}
\end{equation}
This is shown in Fig.\ref{fig:V(n) and parabola}. In this case, the inequality can also be proven by an elementary argument, i.e. one looks for the parabola with a unique intersection with $V(n)$. This then leads to the previous bound. In the non-integer case, the same argument is much more involved and involves much case by case work. A much simpler argument which works in general was suggested by Claude. We first define
\begin{equation}
    f(n) = \frac{\pi^2}{6} n^2 -\kappa\abs{n} =\frac{\pi^2}{6} [(\abs{n} -q)^2-q^2] \ , 
\end{equation}
which is basically just the single-site energy we minimized in eq.\eqref{eq:monopole ground state inequality}. Its absolute minimum is at $\abs{n}=[q]$. Then a few lines of algebra show that in fact 
\begin{equation}
    V(n) = f([q]+n)-f([q]) - \frac{\pi^2}{6} n^2 \ .
\end{equation}
Now, since $f(n)$ has its absolute minimum at $n=[q]$ (it could also have more), one has the inequality
\begin{equation}
    f([q]+n)-f([q]) \geq 0 \ .
\end{equation}
This then immediately implies the lower bound,
\begin{equation}
    V(n) \geq - \frac{\pi^2}{6} n^2 \ , 
\end{equation}
which is what we needed. Note that this bound is especially nice since it is uniform in $q$. The above bound then implies
\begin{equation}
    y_n \leq \exp{\pqty{-\beta c_1 n^2}} \ , \qquad  c_1 = \pi^2(2G(0) -1/6) \ .
\end{equation}
Therefore this argument is valid as long as $c_1 > 0$, which is true for the cubic lattice. 

Then we can rearrange the sum, which is valid due to the absolute summability of the discrete Gaussian, as follows:
\begin{align}
    2\sum_{D =1}^{\infty} K(D) &= \sum_{n,m \in \Z, n\neq m} y_n y_m =\\
    &=\pqty{\sum_{n\in \Z} y_n}^2 - \pqty{\sum_{n\in \Z} y_n^2} \leq \pqty{\sum_{n\in \Z} y_n}^2 - y_0^2
\end{align}
Now let $t = \exp(-\beta c_1)$. Since $n^2 \geq n$ for $n \geq 1$, we have $t^n = \exp(-\beta c_1 n ) \geq \exp(-\beta c_1 n^2 )=y_n$. Further, $y_0=1$. Then we can bound the sum,
\begin{equation}
    \sum_{n\in \Z} y_n = 1 + 2 \sum_{n=1}^\infty y_n \leq 1 + 2 \sum_{n=1}^\infty t^n =\frac{1+t}{1-t} \ ,
\end{equation}
since $t < 1$. Therefore, 
\begin{equation}
    2\sum_{D =1}^{\infty} K(D) \leq \frac{4t}{(1-t)^2} \ .
\end{equation}
Now if we can ensure that 
\begin{equation}
    \frac{4t}{(1-t)^2} < 1 \ .
\end{equation}
then the required inequality is satisfied. In the large $\beta$ regime, this is true as long as
\begin{equation}
    t < 3-2\sqrt{2} \approx 0.17 \ .
\end{equation}
Or, equivalently,
\begin{equation}
    \beta > -\log(3-2\sqrt{2})/c_1 \approx 0.53
\end{equation}
where we used $c_1 \approx 3.34$. Note that this bound is not tight, but it is sufficient for our purposes. This concludes the proof of positivity, and in particular implies that we can define a potential for $\chi$ via
\begin{equation}
    \exp(-U(\chi_x)) = K(0)+2\sum_{D =1}^{\infty} K(D) \cos{\pqty{\chi_x D}} \ .
\end{equation}
From the previous estimates, we have, roughly,
\begin{equation}
    K(D) \approx \exp(-\mathrm{const} \times \beta D^2) \ . 
\end{equation}
To leading order, we therefore get a sine-Gordon model, i.e.
\begin{equation}
    U(\chi_x) \approx -\log\pqty{1+2\frac{K(1)}{K(0)}\cos(\chi_x)} \approx -2\frac{K(1)}{K(0)} \cos(\chi_x)
\end{equation}
Now since we set $V(0)=0$, $K(0)\approx \rho(0,0)=1$ up to exponentially small corrections in $\beta$. On the other hand,
\begin{align}
    K(1) &\approx y_1 y_0+y_{-1}y_0 =\\
    &=e^{-2\pi^2 G(0)\beta} \pqty{e^{-\beta V(-1)} + e^{-\beta V(1)} } \ .
\end{align}
Substituting $V$ and additively normalizing $U$ so it is positive, we find,
\begin{equation}
    U(\chi_x) = 2z(\beta,q) (1-\cos(\chi_x)) \ ,
\end{equation}
where
\begin{equation}
    z(\beta, q) = 2 e^{-2\pi^2G(0)\beta}\begin{cases}
         \cosh\pqty{\frac{\pi^2}{3}\beta ([q]-q)} & q >  1/2\\
         \exp\mu & q < 1/2
    \end{cases}
\end{equation}
which is the final result.

\subsubsection{Sum and difference variables with Pauli-Villars regulator} \label{sec:sum and difference PV}

In this section we derive the expression for the Pauli-Villars kinetic term in terms of sum and difference variables. We then discuss the result of integrating out the difference field. The Pauli-Villars operator which appears in the action was introduced in eq.\eqref{eq:PV split} and is given by
\begin{equation}
    \square_{\mathrm{PV}} = \square + \frac{\square^2}{M^2} \ .
\end{equation}
It is easier to work in momentum space, where the Laplacian is diagonal. For concision, we write
\begin{equation}
    \square(p)=\sum_{\mu=1}^3 \hat{p}_\mu^2 \ , \qquad \hat{p}_\mu =2 \sin(p_\mu/2) \ .
\end{equation}
The Pauli-Villars kinetic term is given by
\begin{equation}
    S_{PV} = (\phi, \square_{PV} \phi) = \int_{\mathrm{BZ}} \frac{d^3 p}{(2\pi)^3} \,\,\abs{\phi_p}^2 \square_{PV}(p) \ .
\end{equation}
Here the Brillouin zone of the cubic lattice is $\mathrm{BZ}=\bqty{-\pi, \pi}^3$, and we have a reality constraint $\phi_{-p}=\phi_p^*$.  We now want to introduce sum and difference variables, but these live on the even sublattice. Therefore their momenta belong to a different Brillouin zone $\widetilde{\mathrm{BZ}}$, which is a subset of the Brillouin zone of the cubic lattice. Its derivation is standard textbook material, and it leads to a description of $\widetilde{\mathrm{BZ}}$ as the original Brillouin zone $\mathrm{BZ}=\bqty{-\pi, \pi}^3$, where one identifies momenta which differ by
\begin{equation}
    Q=(\pi,\pi,\pi) \ .
\end{equation}
Geometrically, it looks like a truncated octahedron inside the cubic Brillouin zone, occupying exactly half its volume. We won't need an explicit description of $\widetilde{\mathrm{BZ}}$ beyond this. Therefore the regulated action can be written as
\begin{equation}
    S_{PV} = \int_{\widetilde{\mathrm{BZ}}} \frac{d^3 p}{(2\pi)^3} \,\, \bqty{\abs{\phi_p}^2 \square_{PV}(p) + \abs{\phi_{p+Q}}^2 \square_{PV}(p+Q)} \ .
\end{equation}
Note that the correct normalization for the momentum integral over $\widetilde{\mathrm{BZ}}$ would be $(2\pi)^3 /2=4\pi^3$ since its volume is half the volume of $\widetilde{\mathrm{BZ}}$. We will need to take this into account at the end. 

The lattice Fourier transform of the field can be written as
\begin{align}
    \phi_p &= \sum_x e^{-ipx} \phi_x =\\
    &= \sum_{x\,\mathrm{even}} \bqty{e^{-ipx} \phi_x +e^{-ip_1} e^{-ipx} \phi_{x+\hat{1}}}\  .
\end{align}
Then we can substitute the sum and difference fields according to eq.\eqref{eq:sum and difference variables}. Note that since they live on the even sublattice, their Fourier transform is defined by summing over even lattice sites. Therefore,
\begin{equation}
    \phi_p = \chi_p(1+e^{-ip_1}) + \xi_p(1-e^{-ip_1}) \ .
\end{equation}
Then since $e^{iQ x} =(-1)^x$, one has $\chi_{p+Q}  = \chi_p$ since the sum runs over even sites only, and similarly for $\xi$. Substituting into the action, some algebra leads to 
\begin{equation}
    S_{PV} = \int_{\widetilde{\mathrm{BZ}}} \frac{d^3 p}{(2\pi)^3} \,\, \Bigg[ A(p)\abs{\chi_p}^2 + B(p) \abs{\xi_p}^2 + \Im \bqty{\chi_p^*  \xi_p C(p)}\Bigg] \ .    
\end{equation}
where
\begin{align}
    A(p) &=(2+2\cos p_1) \square_{\mathrm{PV}}(p) +(2-2\cos p_1) \square_{\mathrm{PV}}(p+Q)  \\
    B(p)&=(2-2\cos p_1) \square_{\mathrm{PV}}(p) +(2+2\cos p_1) \square_{\mathrm{PV}}(p+Q)\\
    C(p)&=-4\sin p_1 \bqty{\square_{\mathrm{PV}}(p) - \square_{\mathrm{PV}}(p+Q)}
\end{align}
Note that at low energies, neglecting for the moment the Pauli-Villars mass, one has
\begin{align}
    A(p) &\approx 16p_1^2+4p_2^2 + 4 p_3^2 \ , \\
    B(p)&\approx 48 -12p_1^2 -4p^2 \ .
\end{align}
Therefore $\chi$ has a massless, yet anisotropic, dispersion relation. This will be fixed by integrating out $\xi$. On the other hand, $\xi$ has a mass of the order of the lattice spacing.

As discussed in the main section, the next step consists in integrating out $\xi$. The calculation is just an exercise in Gaussian integration, but it is complicated by the need to impose the reality constraint on the scalar field. We are interested in computing an observable of the form,
\begin{equation}
    \exp\pqty{i \sum_{x \,\mathrm{even}} \xi_x w_x} = \exp\pqty{i\int_{\widetilde{\mathrm{BZ}}} \frac{d^3 p}{4\pi^3}\,\, \xi_p \widetilde{w}_{-p}} \ .
\end{equation}
Note the factor of $2$ arising from the Fourier transform over the even sublattice. Then we need to compute
\begin{equation}
    E \equiv \int D\xi \, \exp\pqty{-\frac{1}{8\pi^2\beta }S_{PV}} \exp\pqty{i\int_{\widetilde{\mathrm{BZ}}} \frac{d^3 p}{4\pi^3}\,\, \xi_p \widetilde{w}_{-p}} \ .
\end{equation}
After performing the integration and ignoring some prefactors, we find the exact result (no approximations):
\begin{multline}
    E = \exp\Bigg[\int_{\widetilde{\mathrm{BZ}}} \frac{d^3 p}{(2\pi)^3} \Bigg( -\frac{1}{8\pi^2\beta }\pqty{A(p)-\frac{C(p)^2 }{4 B(p)} }\abs{\chi_p}^2+\\ - \frac{8\pi^2\beta }{B(p)} \abs{\widetilde{w}_p}^2 +i\frac{C(p)}{B(p)} \Im (\chi_p \widetilde{w}_p^*)  \Bigg)\Bigg] \ .
\end{multline}
Now one can see that, to leading order,
\begin{equation}
    A(p)-\frac{C(p)^2 }{4 B(p)} \approx 4(p_1^2 + p_2^2 + p_3^2)\  .
\end{equation}
so that the kinetic term for $\chi$ is now isotropic at low energies. This is the first piece of information. The cross-term is instead a higher derivative correction, as in the standard result, and can be neglected. 

The main difference compared to the calculation in the main section involves the $\abs{\omega_p}^2$ term, where $\omega_x = n_x+m_x$. Ignoring the Pauli-Villars cut-off leads to a non-negligible contribution from this term, but here we show it can be neglected. Its coefficient is given by 
\begin{equation}
    0 \leq \frac{8\pi^2 \beta}{B(p)} \leq \frac{8\pi^2 \beta}{\mathrm{min}\,B(p)} \ .
\end{equation}
The absolute minimum of $B$ can be computed and is given by
\begin{equation}
    \mathrm{min} \,B = 16 + \frac{64}{M^2} \ .
\end{equation}
Therefore we have the inequality,
\begin{equation}
    0 \leq \frac{8\pi^2 \beta}{B(p)} \leq \frac{8\pi^2 \beta M^2 }{16 M^2 + 64} \leq \frac{\pi^2 \beta M^2 }{8} \ .
\end{equation}
The key point here, as previously discussed at the beginning of Section \ref{sec:effective theory} and in particular in footnote \ref{foot:gopfmack clarification}, is that $M$ must be chosen so that $\beta M \to 0$ exponentially with $\beta$. Then the corresponding term in the sum is the exponential of an exponentially small contribution and is thus one up to exponentially small corrections in $\beta$. The reason why this argument works is that, if we were to restore the lattice spacing, no factors of $a$ (and in particular, no explicit factors of the momentum) would appear in the $\abs{\widetilde{w}_p}^2$ term. For the other two terms, this argument would not give relevant information.

\subsection{Connection to other models}

In this section, we discuss the connection of the model we consider in this work to other theories. 

\subsubsection{Long range Ising model}\label{sec:long-range ising}

We have seen that for large $\beta$ and sufficiently large $\mu$ the monopoles are arranged antiferromagnetically. This ordered ground state breaks a $\Z_2$ charge conjugation symmetry. It is then natural to expect that for small $\beta$, the symmetry is restored with the two phases separated by a phase transition. 

If we fix $\mu/\beta$ so that the dominant monopole charge is constant, and vary $\beta$ a natural approximation would be to fix 
\begin{equation}
    n_x = [q] s_x \qquad s_x=\pm1 \ .
\end{equation}
In other words, near the phase transition we expect the spin flips to be the dominant excitation. With this substitution, the monopole partition function becomes that of a type of long-range Ising model with energy
\begin{equation}
    \label{eq:long-range Ising energy}
    E_M(n)=2\pi^2 [q]^2 (s, G s) \ .
\end{equation}
This model was studied in \cite{LongRangeIsing}, where it was found that it undergoes a phase transition at some finite value of $\beta$, in the short-range Ising universality class. It may be somewhat surprising that the transition of a long-range model with Coulomb interactions should be the same of a model with short-range interactions, but this is due to a phenomenon which we have encountered several times already in this work. Because the order is antiferromagnetic and the Néel state corresponds to
\begin{equation}
    \vec{p}_0=(\pi,\pi,\pi) \ ,
\end{equation}
one finds,
\begin{equation}
    G(\vec{p}_0+\vec{p})= \frac{1}{12} + \frac{1}{144} \vec{p}^2 + \mathcal{O}(\vec{p}^4)
\end{equation}
In other words, in an effective theory description the order parameter (i.e. the staggered magnetization $M_s$) would have a massive dispersion relation, which makes it short-ranged. This is the same reason why we could integrate out the difference field $\xi$ in Section \ref{sec:effective theory}.  

Results of this model then give us some expectation as to the behavior of our theory in the small or moderate $\beta$ regime. In particular, we expect a phase transition associated with $\Z_2$ symmetry breaking and in the short-range Ising universality class. By inspecting eq.\eqref{eq:long-range Ising energy}, one expects a $q$ dependence of the critical coupling $\beta_c(q)$ given by
\begin{equation}
    \label{eq:transition q dependence prediction}
    [q]^2 \beta_c(q) \sim \mathrm{const}.
\end{equation}
Near the first-order transition point, we should perhaps not take the above equation too seriously. 

\subsubsection{Spin ice}\label{sec:spin ice}

The theory considered in the present work bears some connection with condensed matter theory descriptions for \textquote{spin ice} \cite{Castelnovo2008, Cathelin}. These are materials described by spin systems, where however, in certain regimes, the effective degrees of freedom are monopoles. The effective theory for the monopoles is closely related to the monopole gas that we considered in Section \ref{sec:monopole gas}. More specifically, the relevant model in condensed matter is known as the \textquote{dumbbell model}. In particular, while the monopoles are usually dilute, they can be driven to a dense phase by the application of a magnetic field which polarizes the original spins and effectively lowers the energy cost of monopoles.

Of course there are many differences with our theory. Not only are the observable of interest in condensed matter theory different from the ones we've considered, but the spins in spin ice materials live on the primal pyrochlore lattice, while the monopoles live in its dual, the diamond lattice. Nonetheless, since the diamond lattice is bipartite, some of our results may be of interest also for spin ice.

\subsection{Summary of analytical predictions}\label{sec:analytic summary}

\begin{figure}
    \centering
    \begin{tikzpicture}[>=stealth]
    \fill[blue!12]
        (0,-0.7) --
        (4.5,-0.7) --
        plot[domain=4.5:0.25, samples=150, variable=\x]
            ({\x},{1/\x}) --
        (0,4.1) -- cycle;

    \fill[blue!12]
        (0,-0.7) --
        (4.7,-0.7) --
        (4.7,0.5) --
        (0.0,0.5) -- cycle;

    \draw[->] (0,0) -- (5,0) node[right] {$\beta$};
    \draw[->] (0,-0.7) -- (0,4.2) node[above] {$q$};

    \foreach \y/\lab in {
        0.5/{\frac12},
        1.5/{\frac32},
        2.5/{\frac52},
        3.5/{\frac72}
    } {
        \draw[dashed] ({1/\y},\y) -- (4.7,\y);
        \node[left] at (0,\y) {$\lab$};
    }

    \node at (0.9, 1.8) {$?$};
    \node at (0.67, 2.8) {$?$};
    \node at (0.60, 3.8) {$?$};

    \foreach \n in {1,2,3} {
        \node at (2.5,\n) {$n=\pm \n$};
    }
    \node at (2.5, -0.1) {$n=0$};

    \draw[thick, domain=0.25:2.0, samples=150, variable=\x]
        plot ({\x},{1/\x});
\end{tikzpicture}
    \caption{Sketch of the phase diagram in the $\beta , q$ plane as predicted from our analytical arguments. Here we plot $q$ where $\mu\equiv \frac{\pi^2}{3}q\beta$ for convenience. The shaded region is the dilute phase where, at least at large $\beta$, the dominant contribution comes from $n=0$ monopoles. The dashed lines are first-order transitions beyond which staggered configurations of higher monopole charges dominate. The small $\beta$ phase is analytically connected to the dilute phase at large $\beta$. The shape of the smooth curved line is only indicative and should not be taken seriously, especially near the intersections with the horizontal lines (strictly speaking, the prediction would be a step function as in eq.\eqref{eq:transition q dependence prediction}); we expect that it should be a line of second order transitions in the Ising universality class. The question marks indicate that we do not know whether the corresponding horizontal lines intersect the Ising line, or whether they end at a critical point. }
    \label{fig:phase diagram sketch}
\end{figure}
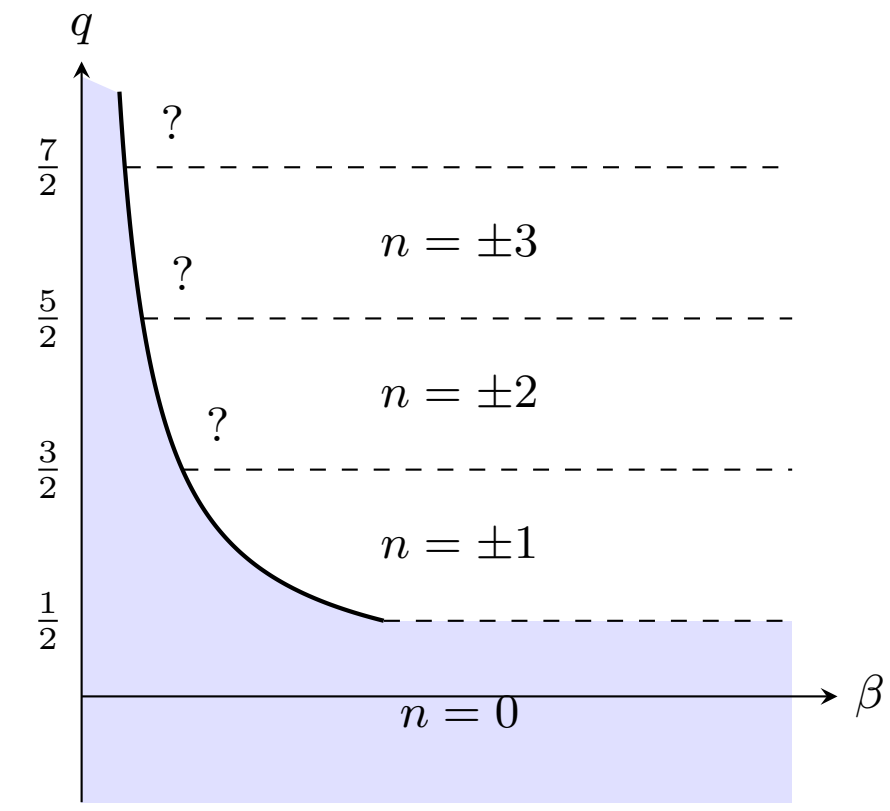

Finally, we collect the results obtained in this section in a sketch of the phase diagram, shown in Fig. \ref{fig:phase diagram sketch}. The phase diagram is best organized in the $\beta- \mu/\beta$ plane. In particular we use $q \in \R$, where $\mu/\beta = q\pi^2/3$, as in the previous sections, due to its direct connection with the ground state. 

At large $\beta$, the dominant monopole charge is given by $[q]$ (we assume $q> 0$) and one has first-order transitions between monopole charges of different values at $q \in \Z_{\geq 0} + \frac12$. Increasing $\beta$ at fixed $q$ leads to a continuum limit which is qualitatively the same as the one of the standard theory. In particular the ratio $m^2 / \sigma \to 0$ as $\beta \to \infty$. We emphasize that this derivation does not apply on the first-order lines $q \in \Z_{\geq 0} + 1/2$, and it is possible that sending $\beta \to \infty$ on this line might lead to different behavior. 

As $\beta$ is decreased at fixed $q>1/2$ we expect instead a second-order transition in the Ising universality class, whose order parameter is the staggered monopole magnetization. It is not clear what is the shape of the line of these phase transitions. One possibility is the step function eq.\eqref{eq:transition q dependence prediction}, but it could also plausibly be a smooth curve. We do not expect the horizontal lines in Fig.\ref{fig:phase diagram sketch} to remain horizontal also at small $\beta$, but we do not have predictions for their shape. Moreover, since the dense phases are not separated by symmetry, it is also possible that the horizontal lines separating dense phases (i.e. $q \in \Z_{\geq 1}+\frac12$) never meet the Ising line and instead end on a critical point (like the liquid-gas line in the phase diagram of water). We do not make predictions on this point and we have added question marks to the figure in order to reflect this unknown. On the other hand, the $q=1/2$ horizontal line should intersect the Ising line and a natural guess would be a critical point in the tricritical Ising universality class.

Note that while we expect the shaded region in Fig.\ref{fig:phase diagram sketch} to form a single phase, the monopoles are dilute and dominated by the neutral configuration only at large $\beta$; in fact, even for the standard theory, the monopole density is divergent as $\beta \to 0$ (this can be easily seen from the monopole gas representation in Section \ref{sec:monopole gas}); for small $\beta$ the monopoles behave more as an almost decoupled \textquote{soup} with wild oscillations. Nonetheless, throughout this phase, they remain disordered in the sense that the staggered magnetization is zero.

Now we briefly consider what would happen, based on our predictions, if, starting in the dilute phase, we increase $\beta$ while varying $\mu$ so as to keep the ratio $\sigma/m^2$ constant. Searching for such a line of constant physics was done numerically in \cite{athenodorou2026differentkindcontinuumlimit}. For large enough $\beta$, from eq.\eqref{eq:string tension prediction} we have
\begin{equation}
    \frac{\sigma}{m^2} \sim \frac{1}{(am)\beta} \ .
\end{equation}
Hence we follow a trajectory with fixed $(am)\beta$, which for convenience we set to a specific value, 
\begin{equation}
    (am)\beta \equiv r \ .
\end{equation}
The behavior of the mass in the dilute phase ($q < 1/2$) is given by eq. \eqref{eq:mass gap pediction}, i.e.
\begin{equation}
    (am)^2 = 8\pi^2 \beta \exp(-2\pi^2G(0)\beta+\mu) \ .
\end{equation}
Hence the above trajectory requires,
\begin{equation}
    \mu(\beta) = 2\pi^2G(0) \beta -\log\pqty{\frac{8\pi^2 \beta^3}{r^2}} \ .
\end{equation}
For ease of comparison, we replace $\mu$ with $q$, obtaining
\begin{equation}
    q(\beta) = 6G(0) -\frac{3}{\pi^2 \beta}\log\pqty{\frac{8\pi^2 \beta^3}{r^2}} \ .
\end{equation}
In the limit $\beta \to \infty$ the second term goes to zero, and we see that
\begin{equation}
    \label{eq:constant ratio trajectory infinite beta}
    q(\beta) \to 6  G(0) \approx 1.52 > 1/2 \ . 
\end{equation}
Therefore at some point the trajectory will hit $q=1/2$ at which point it will undergo a first-order transition and the previous formulas are no longer valid. The ratio approached by $q$ in this argument is a funny combination of lattice quantities, i.e. tracing this through the calculation, 
\begin{equation}
    \frac{G(0)}{2\,\mathrm{min}_p\,\widetilde{G}(p)} \ .
\end{equation}
One may then naturally wonder whether there might exist lattices for which this is less than $1/2$. But this is not possible, as, under minimal assumptions, one can write,
\begin{equation}
    G(0) = \int_{\mathrm{BZ}} \frac{d^3 p}{\abs{\mathrm{BZ}}} \widetilde{G}(p) \ ,
\end{equation}
which is just the average value of $\widetilde{G}(p)$ across the Brillouin zone, and is therefore strictly greater than its minimum,
\begin{equation}
    G(0)>\mathrm{min}_p\,\widetilde{G}(p) \ .
\end{equation}
One could also look at a trajectory with fixed $m^2/\sigma$ starting from $q > 1/2$. A similar argument again shows that $q$ would have to grow or shrink until it hits a first-order transition, where the $\beta \to \infty$ condition is again given by eq.\eqref{eq:constant ratio trajectory infinite beta}, with $q$ replaced by $q-[q]$. Assuming that our formulas also hold for the Wilson action, in which case we use the prefactors from \cite{TepAthen}, and fixing $m/\sqrt{\sigma} = 2.317$ as was done in \cite{athenodorou2026differentkindcontinuumlimit}, the corresponding trajectory would hit the $q=1/2$ phase transition at $\beta_W \approx 3.56$ which is very large and much larger than the values studied in \cite{athenodorou2026differentkindcontinuumlimit}.

Finally, we consider a question posed in \cite{athenodorou2026differentkindcontinuumlimit}. As we have seen in Section \ref{sec:wilson vs villain}, while the Villain and Wilson action are well established to be equivalent for the standard theory, it is not clear if this remains true for $\mu \neq 0$. Based on our analytical results, it would seem that a restriction of the monopole numbers would not make a qualitative difference as long as it includes the dominant monopole charge predicted by the ratio of $\mu/\beta$. Of course, we don't know the equivalent parameter for the Wilson action but it should remain true that no significant difference should arise from Villain as long as $\mu$ is not too large. If instead it is too large, i.e. $\frac{3}{\pi^2}\mu/\beta$ is much larger than the maximum monopole charge (this equation is quantitatively valid only for the Villain action), then fluctuations around the ground states are completely suppressed and the physics should resemble more closely that of the long-range Ising model discussed in Section \ref{sec:long-range ising}. For the Wilson action on the cubic lattice, as we have seen in Section \ref{sec:wilson vs villain}, the maximum monopole charge is $2$; therefore one should be able to access the $[q]=1$ phase (whose excitations are mostly charge $0$ and $2$ monopoles).

\section{Numerical results}\label{sec:numerical}

In this section, we discuss the results of numerical simulations we have performed in order to check some of our analytical predictions. We limit ourselves to a study of the monopole distribution, and do not attempt to compute either the mass or string tension. In more detail, we look for the following:
\begin{enumerate}
    \item Fixing $\beta$ and varying $\mu/\beta$, we look for first-order transitions in the monopole distribution. In particular, at large $\beta$ we expect the staggered magnetization eq.\eqref{eq:staggered magnetization} to jump by one at the points where $q\in \Z_{\geq 0}+1/2$.
    \item Fixing $\mu/\beta$ (for convenience, we do this at $q$ integer), we vary $\beta$ and look for a phase transition. If it exists, our predictions indicate that it should be in the Ising universality class. We also check whether the excitations over the dense background are dilute at large $\beta$ as we expect.
\end{enumerate}
We also measure various other quantities related to the monopole distribution, such as the proportion of monopoles of each charge and the excess density of monopoles over the staggered background. 

We simulate the Villain action eq.\eqref{eq:u(1) partition function} by successive updates of the real-valued and the integer-valued fields. In particular, the real-valued field can be updated via a heat-bath algorithm since its distribution is Gaussian. \footnote{Note that it would be incorrect to take the field modulo $2\pi$, since the action for the gauge field alone is not invariant under $2\pi$ shifts; one would need to vary the integer field at the same time, as in eq.\eqref{eq:integer shift Z gauge invariance}.} The integer field is instead updated via Metropolis steps with $\pm 1$ proposals. Inspired by \cite{athenodorou2026differentkindcontinuumlimit}, we also add \textquote{overrelaxation} steps. In particular, the real-valued field is reflected across the mean of its conditional distribution; this is an exactly microcanonical step and always accepted. On the other hand, the integer field is reflected across the closest integer to the peak value of the conditional distribution of each $m_p$ separately. Since the peak is generically not located at an integer point, this step is not exactly microcanonical, so it requires an accept/reject step. We consider three different initializations for the fields: cold (all fields are zero), hot (all fields are random) and staggered (zero real-valued field and staggered $\pm 1$ monopoles). Concretely, we obtain the $\pm 1$ monopole configuration by setting $m_p=0$ except on plaquettes in the $x,y$ plane, where we then set $m_{12}(x,y,z)=(-1)^{x+y}$ for $z$ odd and zero otherwise. We will discuss the effect of the choice of initialization in more detail later. 

Observables of interest are primarily the staggered monopole magnetization, which is an order parameter for the crystalline order,
\begin{equation}
    \label{eq:staggered magnetization numerical}
    M_s = \frac{1}{V} \sum_{x,y,z} (-1)^{x+y+z} (dm)_{x,y,z} \ ,
\end{equation}
where $V$ is the number of lattice sites and $dm_{x,y,z}$ is the monopole number computed at the cube starting at $(x,y,z)$ in the positive directions. We have also computed the monopole density, which we define as
\begin{equation}
    \label{eq:monopole density}
    \rho = \frac{1}{V} \sum_{x,y,z} \abs{(dm)_{x,y,z}} \ ,
\end{equation}
as well as the excess density of monopoles over the staggered background
\begin{equation}
    \label{eq:excitation density}
    \rho_{\mathrm{ex}} = \frac{1}{V} \sum_{x,y,z} \abs{(dm)_{x,y,z} - (-1)^{x+y+z} [M_s]} \ ,
\end{equation}
where, for each measurement, $[M_s]$ is the closest positive or negative integer to the staggered magnetization $M_s$. The quantity $\rho_{\mathrm{ex}}$ can also be thought of as the density of excitations. Related to this, we have also measured the fraction of cubes with monopoles of charges $\abs{dm}=0,1,2$.

\begin{figure}
    \centering
    \includegraphics[width=0.9\columnwidth]{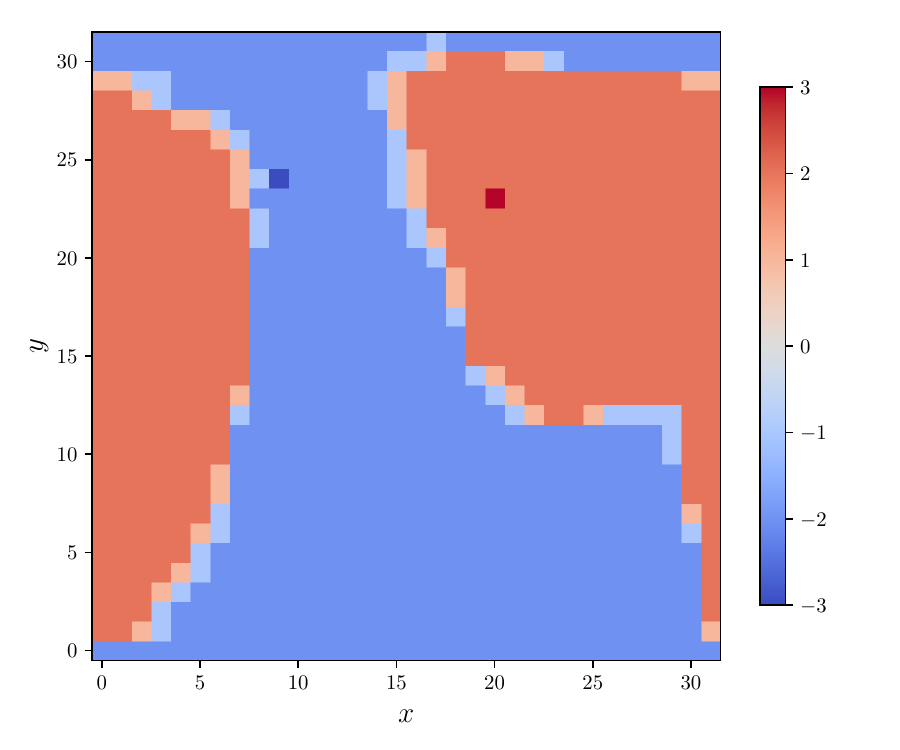}
    \caption{Local staggered monopole magnetization $(-1)^{x+y+z}(dm)_{x,y,z}$ at one fixed $z$ for a single field configuration on a $32^3$ lattice at $\beta=1.5$, $q=2$, obtained from a cold start after $10^5$ sweeps. The algorithm finds the correct staggered charge $\pm 2$, but is unable to eliminate randomly formed domains even after a large number of iterations. This problem occurs generically in the large $\beta, \mu$ region.  }
    \label{fig:domains}
\end{figure}

Before discussing the results, we note that for $\beta \gtrsim 1.6$ and $q \gtrsim 1$ the performance of the algorithm becomes unsatisfactory. In particular, when this occurs, even after $\mathcal{O}(10^5-10^6)$ sweeps the system does not thermalize: runs with different initial conditions (hot or cold) or different initial seeds for the random number generator lead to different results. The fact that numerical issues arise for large $\beta, \mu$ is unsurprising. For a measure of comparison, as shown in Table \ref{tab:wilson villain coupling comparison}, $\beta_V\cong 1.6$ corresponds to a Wilson coupling $\beta_W \approx 2.2$ which is relatively large. However, the source of the difficulty in our case is different than usual: it is caused by the formation of long-lasting domains of the $\Z_2$ symmetry breaking. An example two-dimensional slice of the lattice, showing the domains, can be found in Fig. \ref{fig:domains}. For cold or hot starts, the domains are randomly generated; for smaller $\beta$ in the ordered phase or smaller volumes, they coalesce into a single domain, while for large $\beta$ the domains are instead long-lasting. In particular, this is not an issue of the algorithm \textquote{getting stuck}; the correct staggering with correct charge is found within each domain. We have found that the problem can be solved by initializing the field in a staggered configuration of $\pm 1$ monopoles, which biases the system to form a single domain (in particular, the choice of $\pm 1$ is sufficient to prevent the formation of domains even when the ground state, given our choice of parameter, corresponds to higher charge). Note that this is mostly an issue when exploring the phase diagram at fixed $\beta$ and increasing $q$; the symmetry breaking transition occurs at small $\beta$ and is not affected. We further note that in the ordered phase, tunneling between different vacua of the $\Z_2$ symmetry breaking is strongly suppressed. The lack of tunneling is expected, but appears to be more severe than usual in this model. For example, as we will see later, at $\beta=1.7$ near $q\approx 0.5$ already systems of linear size $L\gtrsim 12$ do not tunnel even after millions of iterations. Perhaps this is due to the effective long-range interactions between the monopoles.

\subsection{Fixed \texorpdfstring{$q$}{q} results}

\begin{figure}
    \centering
    \includegraphics[width=0.9\columnwidth]{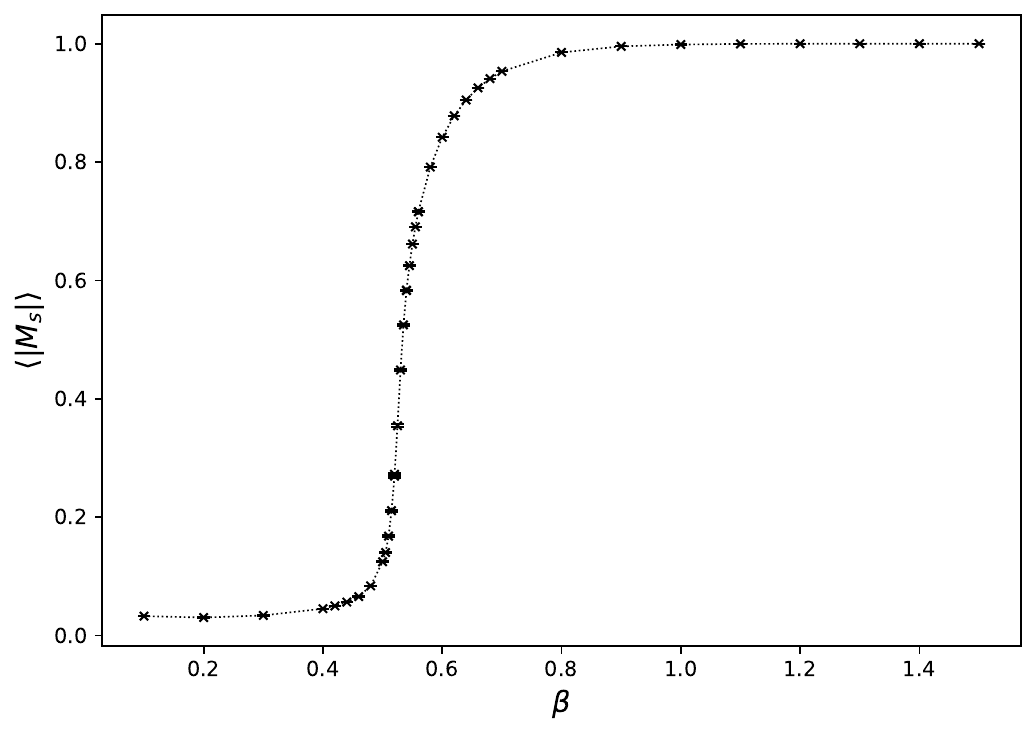}
    \caption{Staggered monopole magnetization vs $\beta$ at fixed $q=1$ on a $16^3$ lattice from a cold start. Errors are shown but are almost too small to be visible. The monopoles are ordered at large $\beta$ and disordered at small $\beta$. The transition appears continuous, consistent with theoretical predictions.}
    \label{fig:Ms vs beta q=1}
\end{figure}

First of all we performed a scan at fixed $q=1$ (i.e. we set $\mu = \frac{\pi^2}{3} \beta$) and varying $\beta$ in the range $[0.1, 1.5]$ on a $16^3$ lattice. The results are shown in Fig.\ref{fig:Ms vs beta q=1}, where we plot the staggered magnetization eq.\eqref{eq:staggered magnetization numerical}. We see that there is a phase transition around $\beta \sim 0.5-0.6$. For small $\beta$ the staggered magnetization is close to zero, while it is close to $1$ for large $\beta$, consistent with our theoretical predictions. Recall that we also predicted that the phase transition should be continuous and in the Ising universality class. While not shown in the figure since the data would simply overlap, we also note that throughout the transition region there is agreement between various initializations (cold, hot or staggered), which is consistent with a continuous transition. We will study this transition in more detail in what follows. 

\begin{figure*}
    \centering
    \includegraphics[width=0.9\columnwidth]{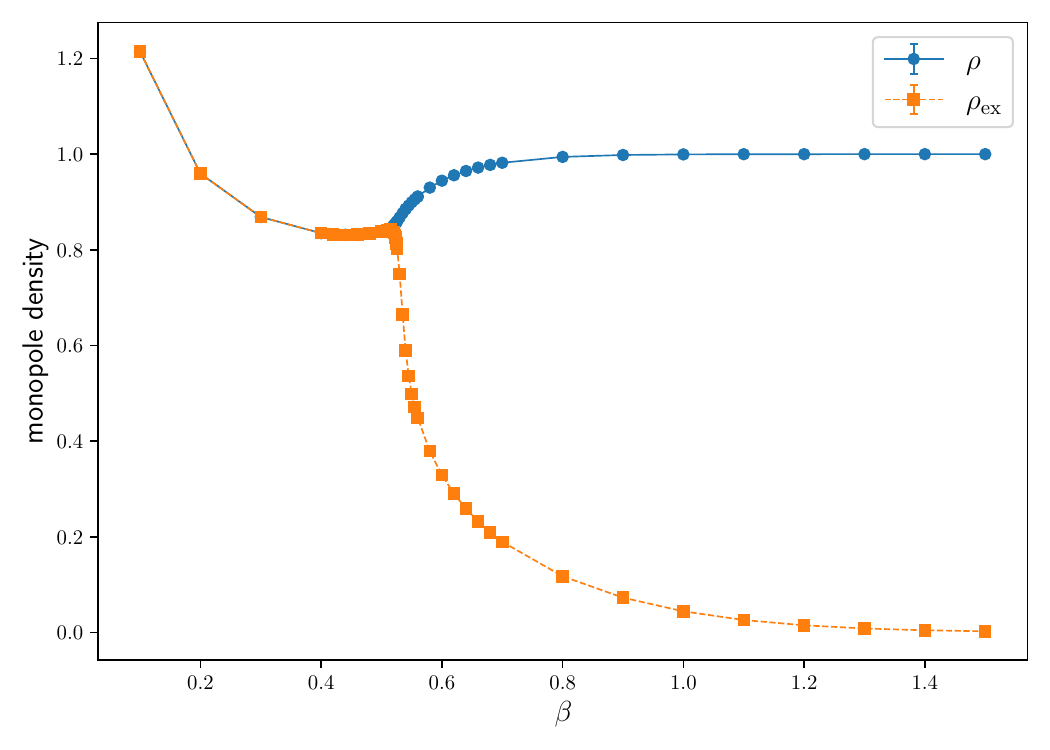}
    \includegraphics[width=0.9\columnwidth]{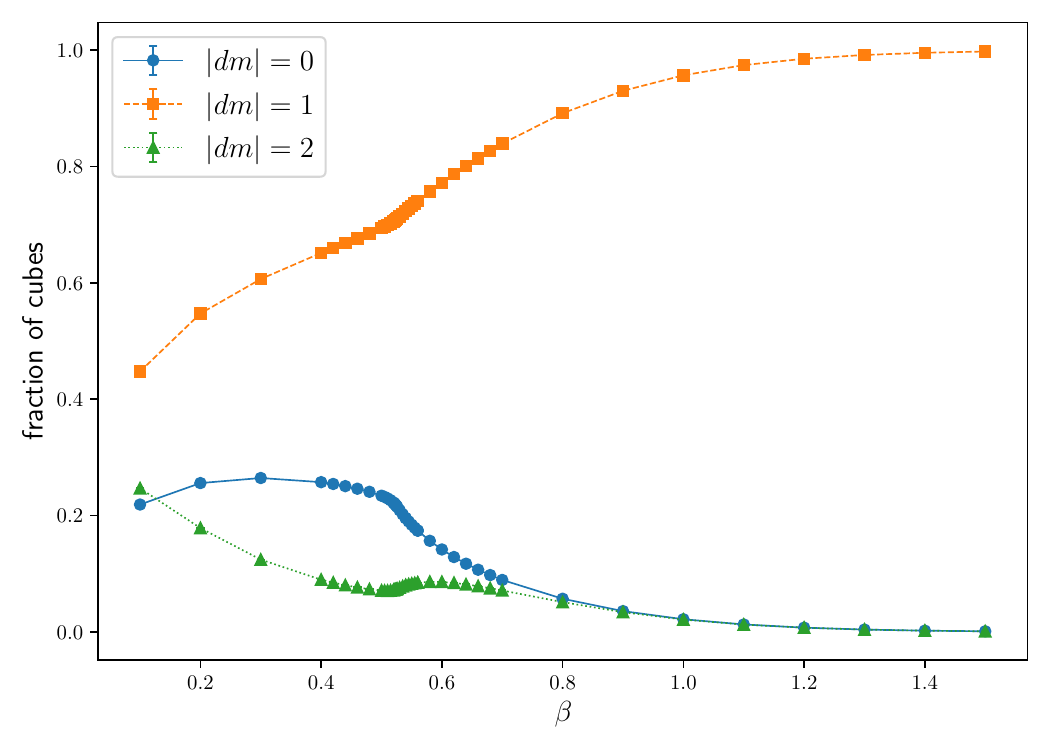}
    \caption{Monopole density and excitation density (\textit{left}) as well as fraction of cubes with given monopole charge (\textit{right}) as a function of $\beta$ at $q=1$, from a cold start. We see that for large enough $\beta$, the dominant charge is $1$ as expected, and the excitations are dilute. In the left plot, $\rho$ and $\rho_{\mathrm{ex}}$ coincide for small $\beta$ since there the staggered magnetization is zero. The discontinuity in $\rho_{\mathrm{ex}}$ is an artifact of its definition.}
    \label{fig:monopole density and charges}
\end{figure*}

With the same parameters (fixed $q=1$ and varying $\beta$) we also plot the density of monopoles eq.\eqref{eq:monopole density} and the excitation density eq.\eqref{eq:excitation density}, as well as the fraction of cubes with monopoles of charges $\abs{dm}=0,1,2$. The results are shown in Fig.\ref{fig:monopole density and charges}. In particular, both plots confirm our expectations that the dominant charge for large $\beta$ is $1$ and that excitations are dilute. Note that the monopole density increases sharply as $\beta \to 0$. This is entirely expected, as even in the standard ($\mu=0$) theory the monopole density is divergent in the limit $\beta \to 0$ (this can be easily seen by setting $\beta = 0$ in the monopole gas representation and computing the relevant observable). For $\mu > 0$ the divergence as $\beta \to 0$ can only become sharper.

\begin{figure*}
    \centering
    \includegraphics[width=0.9\textwidth]{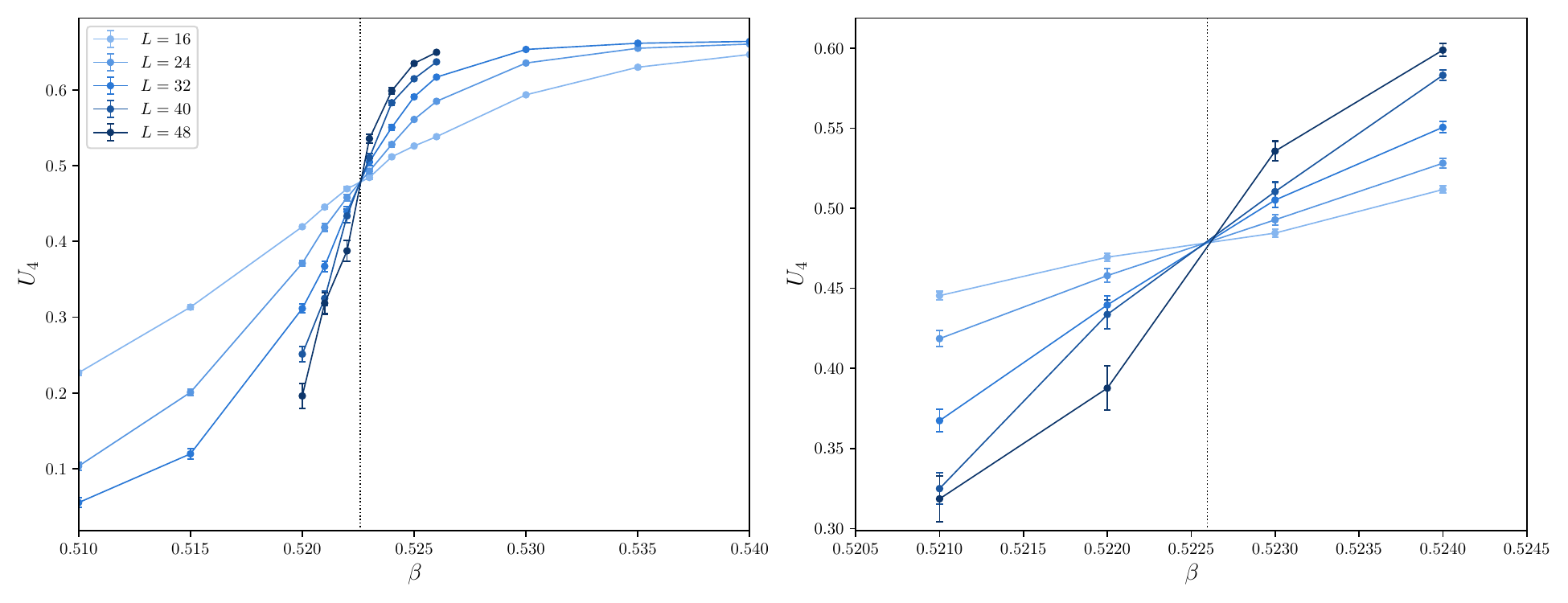}
    \caption{Binder cumulant $U_4$ as defined in eq.\eqref{eq:binder cumulant} as a function of $\beta$ near the phase transition, for increasing lattice size $L$. The figure on the right zooms closer to the location of the transition. The vertical line shows the intersection of the Binder cumulants based on a linear interpolation. }
    \label{fig:binder}
\end{figure*}

We then proceed to study in closer detail the phase transition corresponding to the $\Z_2$ symmetry breaking. A high-precision study of this phase transition would be beyond the scope of this work. We look for some confirmation of our theoretical predictions (continuous transition in the 3D short-range Ising universality class) which can be obtained with moderate numerical effort. We locate the transition by computing the Binder cumulant 
\begin{equation}
    \label{eq:binder cumulant}
    U_4 = 1-\frac{\expval{M_s^4}}{3 \expval{M_s^2}^2} \ .
\end{equation}
In particular, we use isotropic cubic lattices $L^3$ (periodic boundary conditions) at varying $L$, and identify the location of the transition by the crossing of the cumulants at different lattice sizes. The numerical data is shown in Fig. \ref{fig:binder}. The Binder cumulant is well-sampled as, for these values of $\beta$, we find that tunneling occurs relatively frequently in the ordered phase even on the largest lattices we considered. In particular, for each $L$ each curve is monotone and there is a common point of intersection for all $L$, which supports the presence of a continuous transition.  For a first-order transition one would have expected the curves to show a minimum, which becomes more marked with increasing $L$ (see e.g \cite{SenSandvik, Vollmayr1993}). Of course we cannot exclude that the minimum would only show up for larger $L$, which would imply a weak first-order transition.

\begin{figure}
    \centering
    \includegraphics[width=0.9\columnwidth]{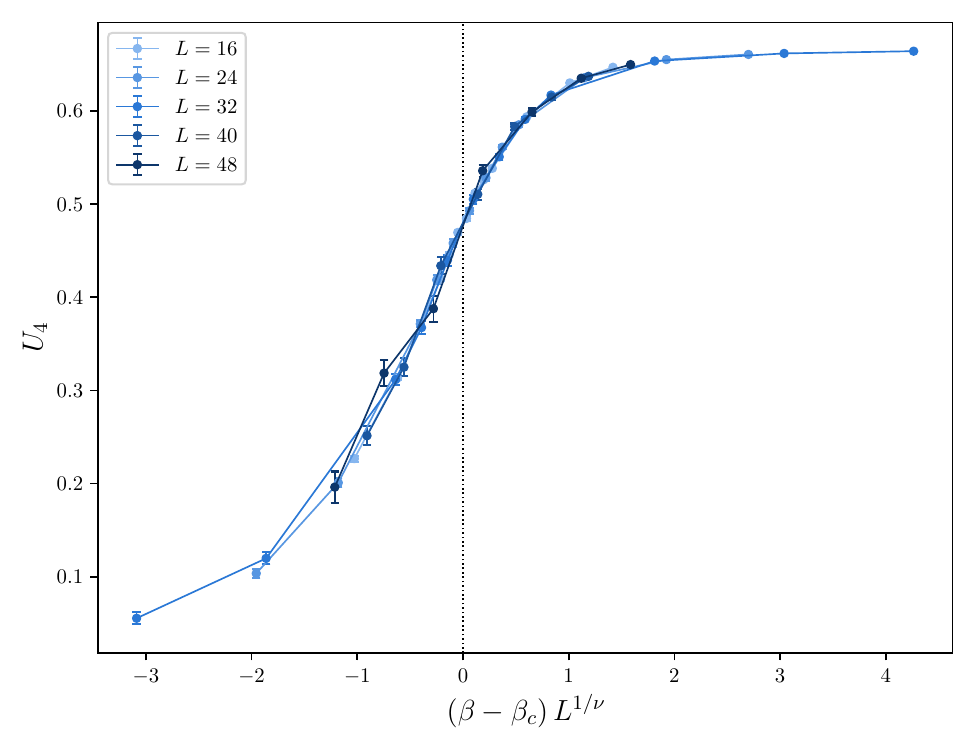}
    \caption{Binder cumulant $U_4$ as a function of $(\beta-\beta_c)L^{1/\nu}$. The value of the critical exponent $\nu$ for the 3D Ising class was taken from \cite{Kos:2016ysd} and $\beta_c$ was fixed to the central value in eq.\eqref{eq:beta_c}. This is the same data as in Fig.\ref{fig:binder}, but plotted on a different $x$-axis; the various curves collapse onto one. }
    \label{fig:binder collapse}
\end{figure}

In order to locate the transition we interpolate the values of the Binder cumulant (either linearly or quadratically), compute the intersections of each pair of lines and report the mean as a final value. This has both a statistical error as well as a systematic error measured from the spread of the values obtained from each pair. The systematic error due to the choice of interpolation turned out to be negligible, and we did not observe drift with respect to $L$. This leads to an estimate for the location of the transition,
\begin{equation}
    \label{eq:beta_c}
    \beta_c = 0.52260(5)(11) \ ,
\end{equation}
with the first error being purely statistical and second being the spread of values. The value of the Binder cumulant is also associated to a universality class and correspondingly we find
\begin{equation}
    \label{eq:measured Binder cumulant}
    U_4^* = 0.4787(29)(75) \ .
\end{equation}
Although less accurate, this seems compatible with the value for the 3D Ising class (isotropic periodic lattice), \cite{Hasenbusch:1998gh}
\begin{equation}
    U_4^* \lvert_{\mathrm{Ising}} = 0.4657(1) \ .
\end{equation}
Near a second-order phase transition, scaling arguments predict that the Binder cumulant $U_4$ should be a function of the combination $(\beta-\beta_c)L^{1/\nu}$. This means that, when plotted in terms of this variable, the various curves in Fig.\ref{fig:binder} should all collapse onto a single curve.  We show this plot in Fig.\ref{fig:binder collapse}, using the published value of the 3D Ising critical exponent $\nu$ from \cite{Kos:2016ysd} and the central value of $\beta_c$ in eq.\eqref{eq:beta_c}. We see that indeed the various curves fall onto one. This provides a check on the determination of the location of the transition and is further evidence that the transition is compatible with the Ising class. 

A more precise determination of the universality class would involve computing critical exponents. Doing so accurately requires very significant numerical effort and would be beyond the scope of this work. A more compelling picture would arise if we could exclude other candidate universality classes. Since the transition is associated with the breaking of a $\Z_2$ symmetry, known alternatives are scarce. It would be reasonable to try to exclude long-range Ising universality classes; but it does not seem these have been much studied in three dimensions, with the exception of the already mentioned \cite{LongRangeIsing} which we discussed in Section \ref{sec:long-range ising}; this model is closely related to ours and it finds short-range Ising exponents. Overall, although more effort would be needed to pin this down beyond reasonable doubt, the numerical data we have obtained are consistent with the theoretical picture of a continuous transition in the short-range Ising class. 

\begin{figure}
    \centering
    \includegraphics[width=0.9\columnwidth]{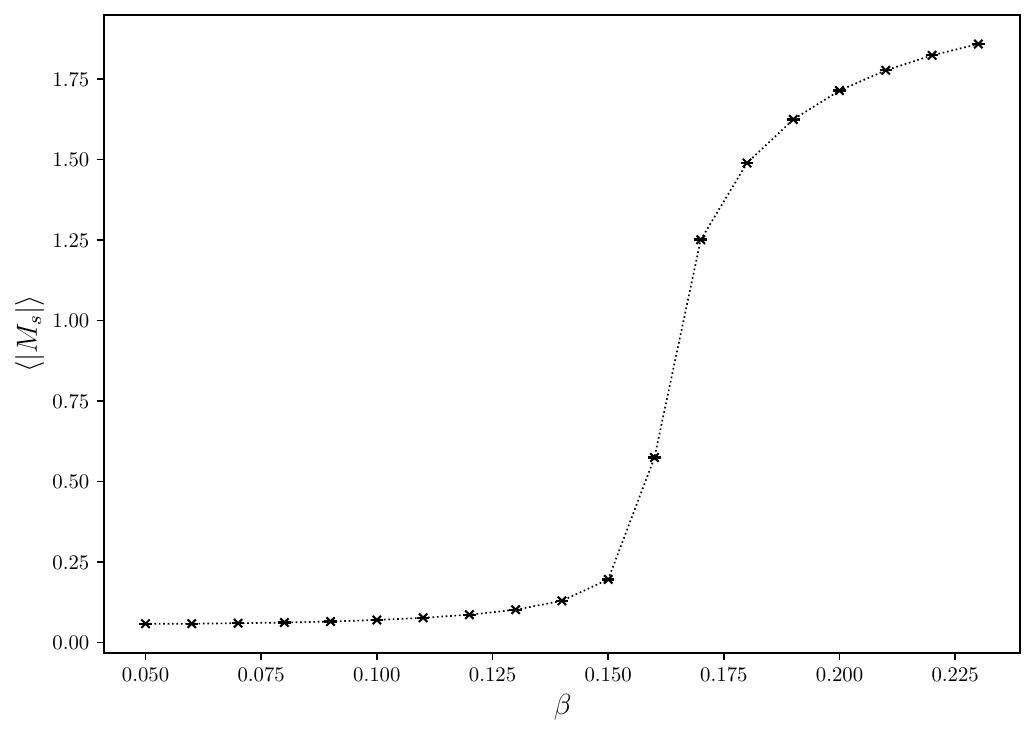}
    \caption{Staggered monopole magnetization vs $\beta$ at fixed $q=2$ on a $16^3$ lattice from a cold start. Errors are shown but are almost too small to be visible. There is a phase transition around $\beta \approx 0.16-0.18$. At large $\beta$, the magnetization approaches the correct value of $2$. }
    \label{fig:Ms vs beta q=2}
\end{figure}

Finally, we briefly consider the phase diagram as $\beta$ is varied at fixed $q > 1$. Recall the prediction from Section \ref{sec:long-range ising} that the $q$ dependence of the critical coupling $\beta_c(q)$ should be given by
\begin{equation}
    [q]^2 \beta_c(q) \sim \mathrm{const}.
\end{equation}
This in particular would lead to the expectation that
\begin{equation}
    \beta_c(2) \approx \frac{\beta_c(1)}{4} \approx 0.131 \ .
\end{equation}
We briefly test this prediction by scanning the region $\beta \in[0.05, 0.23]$ at $q=2$. The results of this scan are shown in Fig. \ref{fig:Ms vs beta q=2} and are compatible with a phase transition around $\beta_c(2) \approx 0.16-0.18$. This is in qualitative agreement with the prediction that the transition should occur at smaller $\beta$, although the quantitative value we find is a bit too large. We have not tested further whether the $\beta_c(q)$ dependence is a step function, as predicted by the same argument, or it is in fact smooth.

\subsection{Fixed \texorpdfstring{$\beta$}{beta} results}

\begin{figure}
    \centering
    \includegraphics[width=0.9\columnwidth]{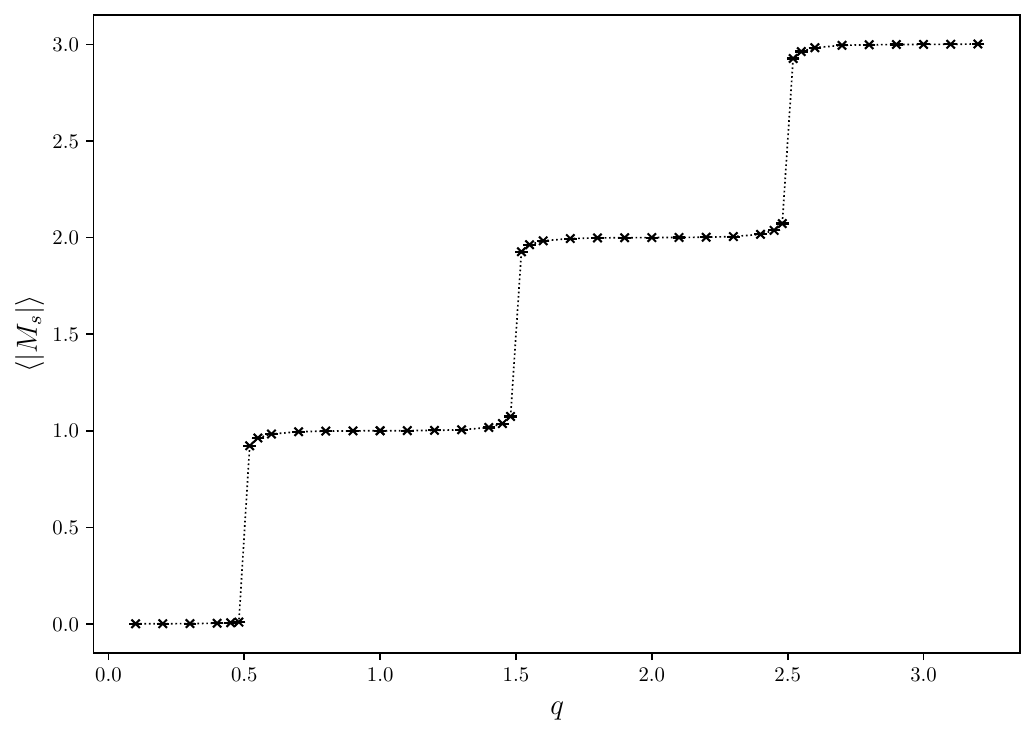}
    \caption{Staggered monopole magnetization vs $q$ at fixed $\beta=1.7$ on a $16^3$ lattice from a staggered $\pm 1$ start. Errors are shown but are almost too small to be visible. The dominant monopole charge jumps abruptly at each $q \in \Z_{\geq 0} +\frac12$ as predicted by the theory. The points closest to the predicted location of the transition, $k+1/2$, for $k=0,1,2,\ldots$, were taken at $q=k+1/2 \pm 0.02$.}
    \label{fig:Ms vs q beta=1.7}
\end{figure}

We then performed a scan at fixed $\beta=1.7$ and varying $q \in [0.1, 3.2]$. Our prediction for the staggered magnetization is that $\expval{|M_s|} = [q]$. Due to the issue mentioned at the beginning of this section, we start from a fully staggered configuration of $\pm 1$ monopoles. The results are plotted in Fig. \ref{fig:Ms vs q beta=1.7} which is consistent with our predictions, and in particular $\expval{M_s} = [q]$ on the plateaus as expected, with jumps at $q \in \Z_{\geq0} +\frac12$. The closest points to the predicted location of the transition, $k+1/2$, for $k=0,1,2,\ldots$, were taken at $q=k+1/2 \pm 0.02$, and they clearly lie on one side or the other. This is evidence of a transition near these points.

\begin{figure}
    \centering
    \includegraphics[width=0.9\columnwidth]{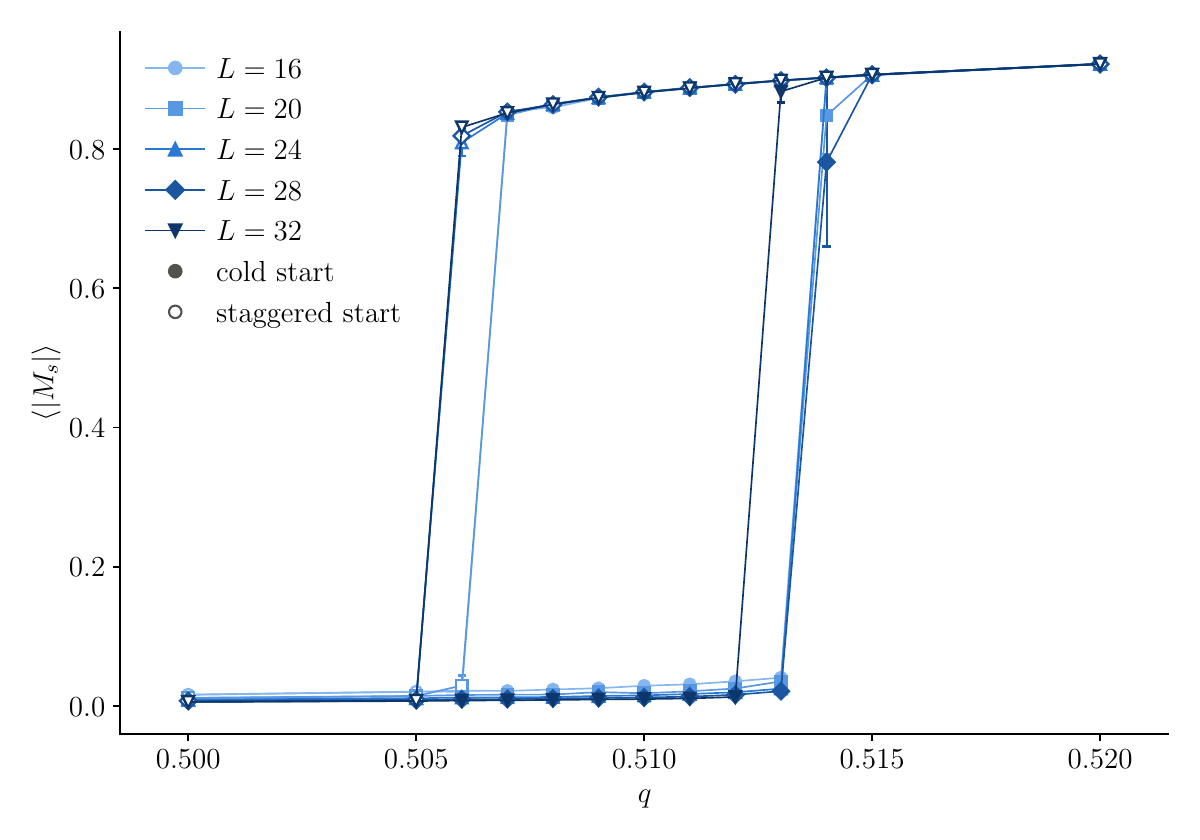}
    \caption{Staggered monopole magnetization vs $q$ at fixed $\beta=1.7$ near $q\approx 0.5$ for various lattice sizes and either cold or staggered start. The various runs agree outside the central range, where they instead disagree and are either ordered or disordered. }
    \label{fig:metastability}
\end{figure}

We then look for evidence that the transitions at approximately $q \in \Z_{\geq 0} + \frac12$ are first-order as predicted. In particular, we focus on $q=1/2$. Strictly speaking, our analytical results predict that the transition is first-order only at $\beta \to \infty$; as $\beta$ is decreased, we expect the $q=1/2$ line to weaken until it intersects the Ising line. The intersection point is likely a critical point in the tricritical Ising universality class. The location of the transition $q_c(\beta) \approx 0.5$ may also change with $\beta$. An initial indication of a first-order transition is the presence of metastability near the transition point, as shown in Fig.\ref{fig:metastability}. This is indicative of a first-order transition in the range $q\approx 0.505-0.513$. 

\begin{table}[]
    \centering
    \begin{tabular}{c|c|c|c|c}
        $L$ & no. sweeps & no. events & sweeps per event\\
        \hline
         $4$ & $5\cdot 10^6$ & $896$ & $\approx 6 \cdot 10^3$  \\
         $6$ & $10^7$ & $315$ & $\approx 3 \cdot 10^4$ \\
         $8$ & $10^8$ & $265$ & $\approx 4\cdot 10^5$\\
    \end{tabular}
    \caption{Number of tunneling events between the dilute and dense phases at $\beta=1.7$ on lattices of size $L^3$ for various $L$ at the approximate coexistence point for each $L$ (see Fig.\ref{fig:coexistence}). A tunneling event is defined to occur when the monopole density crosses from $\rho<0.4$ to $\rho>0.7$, i.e. compare with Fig.\ref{fig:coexistence}.}
    \label{tab:tunneling}
\end{table}

\begin{figure*}
    \centering
    \includegraphics[width=0.9\textwidth]{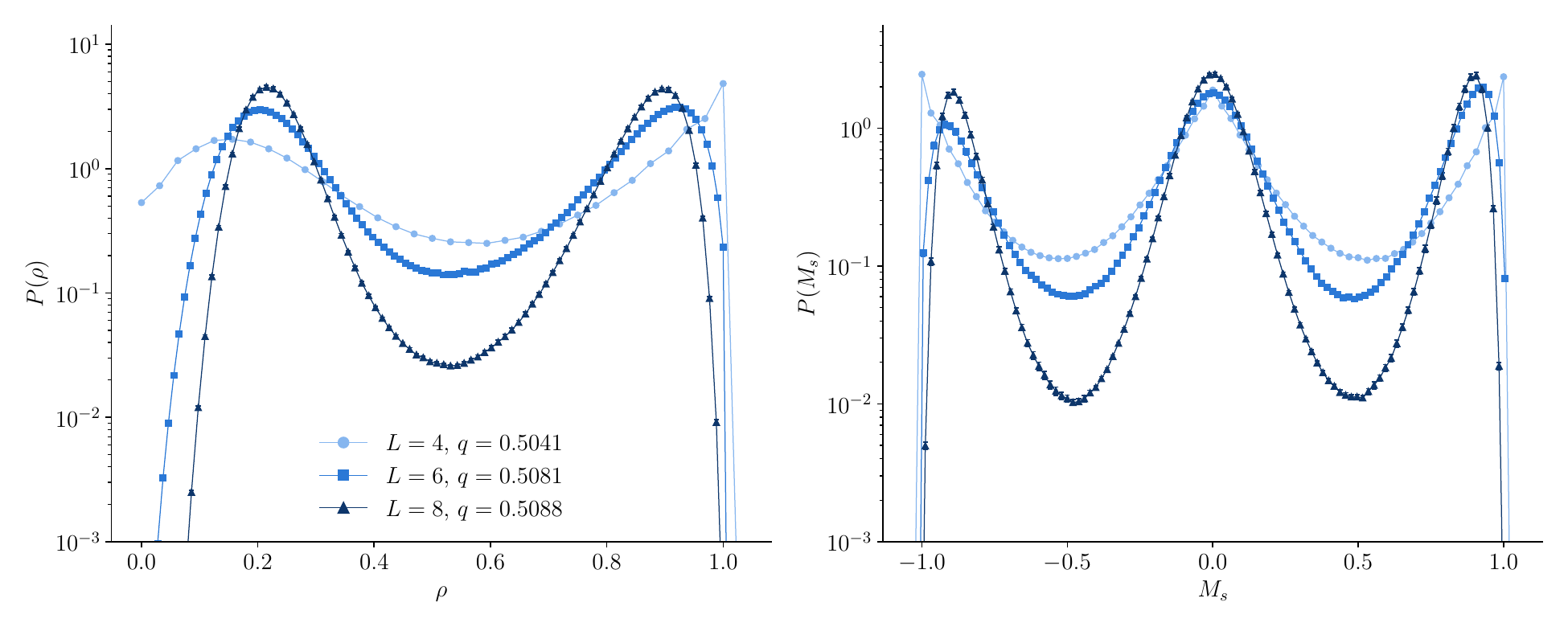}
    \caption{Histogram of the monopole density (\textit{left}) and of the staggered magnetization (\textit{right}) for the three lattice sizes $L=4,6,8$ at $\beta=1.7$ with $q$ set at the approximate equal-weight point for each $L$. The monopole density shows coexistence of the dilute ($\rho \approx 0.2$) and dense ($\rho \approx 0.9$) phases, while the staggered magnetization shows coexistence of the disordered phase ($M_s\approx 0$) with the two broken-symmetry ground states ($M_s\approx \pm 1$). In both cases the peaks become sharper with $L$.}
    \label{fig:coexistence}
\end{figure*}

We then looked for further evidence, in particular histograms of order parameters and the Binder cumulant. Sampling near a first-order transition is complicated by the need to tunnel between different phases in order to sample the coexisting phases. As remarked at the end of the introduction to this section, tunneling is strongly suppressed in this model and we are therefore limited to small lattices. In particular, we performed simulations at $L=4,6,8$, for which we report tunneling rates in Table \ref{tab:tunneling}. At a first-order transition, one expects coexistence of different phases. If they are well-sampled and the transition is first-order, histograms of order parameters should show distinctive peaks which become sharper with the lattice size. A plot of the histogram of the monopole density eq.\eqref{eq:monopole density} and the staggered magnetization eq.\eqref{eq:staggered magnetization numerical} for the three lattice sizes is shown in Fig. \ref{fig:coexistence}. These are shown at the value of $q$ which corresponds to the approximate equal-weight point for the peaks of the distribution for each $L$.

The tunneling time $\tau$ (i.e. the number of sweeps per tunneling event) we report in Table \ref{tab:tunneling} is also in very good agreement with the expected form at a first order transition in 3D, that is $\tau \propto \exp(2\hat{\sigma} L^2)$ (see e.g. \cite{BergNeuhaus}) and with our data we find $\hat{\sigma}\approx 0.044$. This form is due to the fact that for tunneling to occur the system must go through a mixed configuration with two spatial two-dimensional interfaces (since we have periodic boundary conditions). Of course the significance of the agreement is limited by the fact that we only have three points.

\begin{figure}
    \centering
    \includegraphics[width=0.9\columnwidth]{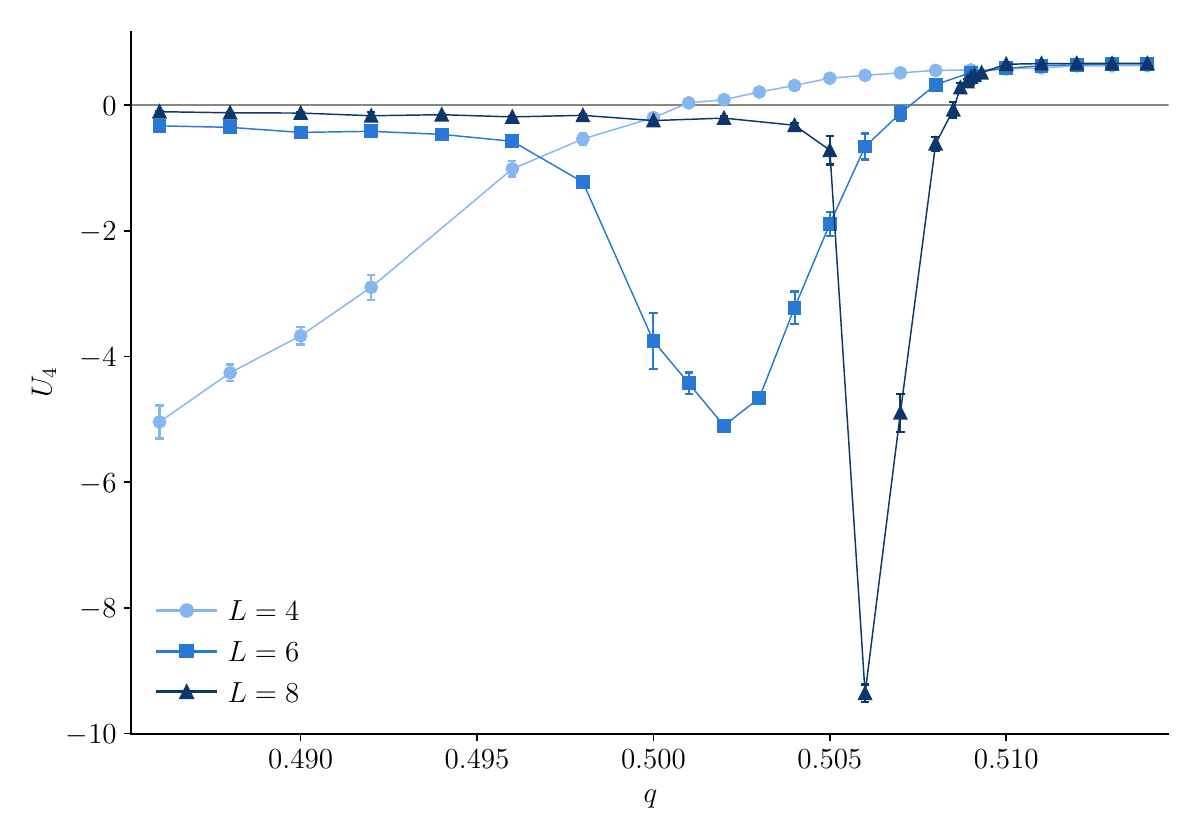}
    \caption{Binder cumulant $U_4$ as defined in eq.\eqref{eq:binder cumulant} as a function of $q$ near the phase transition at $q\approx 0.5$ and fixed $\beta=1.7$, for increasing lattice size $L$. The dip for $L=4$ lies at values of $q$ below those shown. The Binder cumulant shows a dip at a negative value which increases with volume, an indication of a first-order transition.}
    \label{fig:binder first}
\end{figure}

We also computed the Binder cumulant across the transition, at the same values of $L$. The results of the simulations are shown in Fig.\ref{fig:binder first}.  In this case, we see that the Binder cumulant takes the characteristic shape of a first-order transition, with a negative dip near the transition point which deepens with increasing volume (see e.g \cite{SenSandvik, Vollmayr1993}).   This should be compared with the monotone shape we saw at the continuous transition, i.e. Fig.\ref{fig:binder} in the previous section. 

Overall, all the evidence we have considered points towards a first-order transition at $q \approx 0.51$ near the predicted point $q=0.5$. It should be mentioned that some of the evidence such as histograms and the Binder cumulant are obtained on small lattices only and, looking at the numerical data alone, this would leave open the possibility of a different behavior at larger lattices.

\begin{figure}
    \centering
    \includegraphics[width=0.9\columnwidth]{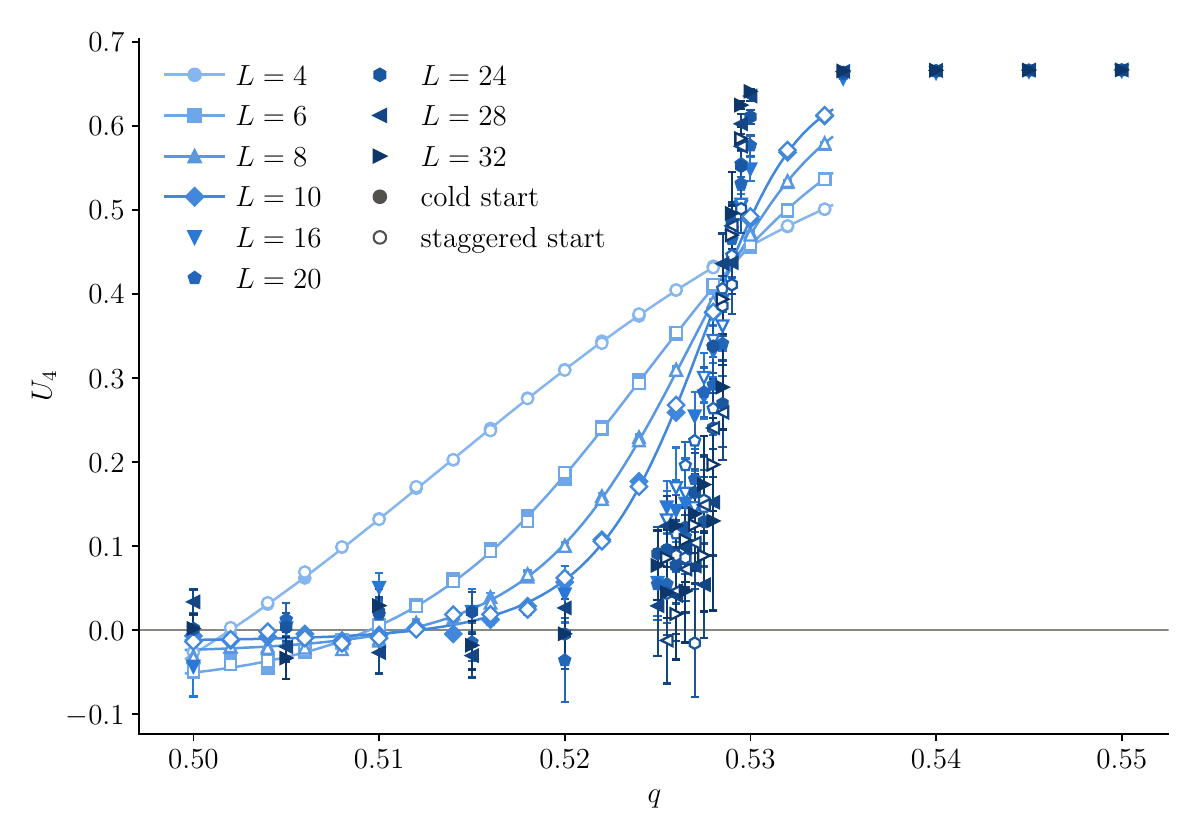}
    \caption{Binder cumulant $U_4$ as defined in eq.\eqref{eq:binder cumulant} as a function of $q$ near the phase transition at $q\approx 0.5$ and fixed $\beta=1.3$, for increasing lattice size $L$ and either staggered or cold start. The curves are monotone and they intersect at a single point, behavior consistent with either a continuous or a weakly first-order transition. }
    \label{fig:binder 13}
\end{figure}

We also briefly report on simulations at smaller values of $\beta$. While our results at $\beta=1.5$ are ambiguous and therefore not shown, at $\beta=1.3$ our data is consistent with either a continuous transition or a weak first-order transition at $q_c \approx 0.528$. In particular, there is no sign of metastability and the Binder cumulant shows the expected behavior at a continuous transition. The Binder cumulant is shown in Fig. \ref{fig:binder 13}. The weakening of the transition with decreasing $\beta$ is expected, but it is not clear whether the transition we found is continuous and already belongs to the Ising line (i.e. the solid curve in Fig.\ref{fig:phase diagram sketch}) or whether it is first-order. More work would be required to determine where exactly the tricritical point lies.

\subsection{Summary of numerical results}

Our numerical results are generally consistent with the theoretical predictions summarized in Section \ref{sec:analytic summary}. In particular and most importantly, we have checked the existence of dense crystalline phases where the monopoles arrange themselves antiferromagnetically, as signaled by the staggered magnetization order parameter. We have also checked that the excitations on top of the dense background are dilute for large $\beta$, which is a key requirement for the applicability of the effective theory we derived. We have also determined the presence of a phase transition as $\beta$ is varied at fixed $q$.

We have also obtained some results on the order and nature of the phase transitions, although more work would be required in order to pin this down beyond reasonable doubt. This would require significant computing time and/or better algorithms. In particular, the transition at fixed $q=1$ around $\beta\approx 0.52$ appears to be continuous, and likely in the Ising universality class. As $q$ is varied around $q\approx 0.5$ at $\beta=1.7$ we found evidence of a first-order transition, which weakens significantly as $\beta$ is lowered.

We have not checked some other aspects of the phase diagram; in particular, we have not determined the shape of the line of Ising transitions, other than the fact that critical value $\beta_c(q)$ seems to decrease with $q$; we have also not determined whether the coexistence lines end on a critical point or whether they intersect the Ising line. 

\section{Conclusions}
\label{sec:conclusions}

We have studied a version of $\U(1)$ lattice gauge theory in three spacetime dimensions, with the addition of a monopole enhancement term controlled by a chemical potential $\mu$. We were able to derive several theoretical predictions for the phase diagram of the model, some of which we also checked by simulating the theory numerically. We were able to show that the continuum limit for $\mu \neq 0$ is qualitatively the same as the one for the standard theory, and in particular the ratio $m^2 / \sigma \to 0$ also in this case. This also answers a question posed in \cite{athenodorou2026differentkindcontinuumlimit}.
The monopole-enhanced theory shows very interesting behavior related to monopoles. For sufficiently large $\mu$ and $\beta$, they form a dense crystal phase, which breaks both charge conjugation and single-site translations. Nonetheless, the excitations on top of the ground state remain dilute, which allows the derivation of an effective theory. At sufficiently large fixed $\mu/\beta$, we also find a transition in the Ising universality class as $\beta$ is varied.

Several aspects of the phase diagram remain underexplored. For example, we have not studied in detail the small $\beta$, large $\mu$ region and in particular we have not determined the shape of the line of Ising transitions. We have also not explored what happens to the first-order transitions associated with coexistence of monopoles of different charges at small $\beta$, i.e. whether they intersect the Ising line or whether they end at a different critical point. The Ising line meets the $q=1/2$ coexistence line at what we conjecture to be a critical point in the tricritical Ising class, but we have not verified this.
It would also be interesting to study this model on a non-bipartite lattice, where the monopole ground state is frustrated. In that case, it would seem that the physics would be dominated by competition between the ground states and it seems difficult to obtain analytical predictions. Finally, it is important to point out that our analytical derivation of an effective theory does not apply along the coexistence lines between different monopole backgrounds. It is possible that increasing $\beta$ along these lines might lead to qualitatively different behavior. 

The main lesson of this work is perhaps that it is sometimes possible to analytically treat dense gases of topological objects, as long as the excitations on top of the dense background are dilute. Our results may also be of interest for quantum simulation, where the $2+1$ dimensional $\U(1)$ gauge theory is often used as a benchmark, as well as for condensed matter physics, due to the connections with spin ice outlined in Section \ref{sec:spin ice}.

As for the question of whether the mass to string tension ratio can be made finite in the continuum limit, our answer is essentially negative. It would be interesting to consider alternative formulations which might allow this, but, in light of the discussion from Section \ref{sec:monopoles}, it is not even clear what would be the target continuum QFT.

This work can be expanded in several directions. One obvious extension would be in four dimensions. In that case, one can perhaps still obtain results about the monopole distribution via the monopole gas representation. We would expect monopole enhancement to enlarge the confining phase, but it is not clear whether it would allow a confining continuum limit, which is not possible for the standard theory. Another interesting extension would be for $\Z_N$ gauge theories in three or four spacetime dimensions. Again we would expect an enlargement of the confining phase and perhaps a better approach to $\U(1)$ in three dimensions.

\section*{Acknowledgments}

The author thanks Claudio Bonati, Michele Caselle and Lorenzo Verzichelli for useful comments. The author is supported by the Simons Foundation grant 994300 (Simons Collaboration on Confinement and QCD Strings). Support from the SFT Scientific Initiative of the Italian Nuclear Physics Institute (INFN) is also acknowledged. The numerical simulations were performed on CINECA computers. The author also benefited from access to Claude via Anthropic's Claude Team for Scientists program.

AI usage statement: the author benefited from AI use (ChatGPT free version, Claude Sonnet and Opus) in the form of conversations on the physics topics as well as in the form of feedback on the manuscript. Claude was also used as a coding assistant. The manuscript was entirely written by the human author. The \texttt{tikz} drawings were drafted by ChatGPT (free version) and modified by the author. 

\bibliography{biblio}

\appendix

\section{Lattice differential forms}\label{sec:lattice differential forms}

In this appendix, we review the theory of lattice differential forms which we used throughout the main text. We work on an arbitrary lattice. We will try to keep the discussion as simple as possible and avoid mathematical subtleties. Lattice differential forms mimic the theory of continuum differential forms and are closely related to several topics in algebraic topology. A physics-friendly introduction to both topics can be found in \cite{nakahara2003geometry}. We also point the reader to a useful review of lattice differential forms from the point of view of the finite-element method \cite{Teixeira:2013kta}.

A lattice in lattice field theory can be seen as a cell complex (for a discussion of lattices as cell complexes aimed at a lattice field theory audience, see \cite{DRUHL1982225}). Concretely, this means that an $n$-dimensional lattice is made of \textquote{cells} of dimension $k=0,1,2,\ldots, n$ which we refer to as $k$-cells. In this language, $0$-cells are lattice sites, $1$-cells are lattice links, $2$-cells are lattice plaquettes and so on. Cells carry an orientation, and a key property is that the boundary of a $k$-cell is a formal sum of oriented $(k-1)$-cells. For example, the boundary of a lattice link is made of two sites with opposite orientation; for a cubic lattice, the boundary of a plaquette is made of four oriented links. This is just a restatement of usual lattice field theory language. Denote with $C_k$ the set of lattice $k$-cells. Then a lattice $k$-form $\alpha_k$ is a map
\begin{equation}
    \alpha_k : C_k \to A\  ,
\end{equation}
where $A$ is an Abelian group, typically $\R, \Z, \C, \U(1)$ or $\Z_N$. Concretely, this means that, say, a $\U(1)$-valued lattice $0$-form is just a $\U(1)$ scalar field or equivalently an assignment of a $\U(1)$ variable to each lattice site; an $\R$-valued lattice $2$-form assigns to each plaquette a real number, and so on. This must be compatible with the orientation, i.e. reversing the orientation of the cell should lead to inversion in the group, 
\begin{equation}
    \alpha(-c_k) = -\alpha(c_k) \ ,
\end{equation}
where $-c_k$ is the $k$-cell $c_k$ with orientation reversed and the minus on the right-hand side refers to inversion in $A$. Call the set of lattice $k$-forms $\Omega^k$. In the mathematical literature, formal linear combinations of cells are called \textit{chains} and lattice differential forms are called \textit{cochains}.

A key operation on lattice differential forms is the exterior derivative $d$, which maps
\begin{equation}
    d : \Omega^k \to \Omega^{k+1} \ ,
\end{equation}
i.e. it increases the form rank. In $n$ dimensions, only cells up to $k=n$ exist and therefore the exterior derivative of an $n$-form is set to zero. The definition of the exterior derivative $d$ is as follows. Given a $k$-form $\alpha$, we know the value of $\alpha(c_k)$ on each $k$-cell $c_k$. Then $d\alpha$ is a $(k+1)$-form, for which we must therefore specify its value on each $(k+1)$-cell $c_{k+1}$. This is done via the rule,
\begin{equation}
    \label{eq:exterior derivative definition}
    d\alpha(c_{k+1})=\sum_{c_k \in \partial c_{k+1}} \alpha(c_k) \ ,
\end{equation}
where we have used additive notation for the Abelian group $A$. In other words, to specify the value of $d\alpha$ on a $k+1$ cell $c_{k+1}$ we consider its boundary $\partial c_{k+1}$, which consists of oriented $k$-cells, and we sum the value of $\alpha$ on these cells taking orientation (and therefore $\pm$ signs) into account. For example, consider a zero form $\alpha$, which is defined on lattice sites. Then $d\alpha$ is a $1$-form, defined on lattice links. Given a link $l: x\to y$, we see that its boundary $\partial l$ consists of two sites, $x$ (negative orientation) and $y$ (positive orientation). Thus,
\begin{equation}
    d\alpha_l = \alpha_y-\alpha_x \ .
\end{equation}
If instead $\alpha$ was a $1$-form, i.e. it takes values on links, then $d\alpha$ takes values on plaquettes. A key property of the exterior derivative $d$ is that $d^2=0$. This is due to the fact that applying eq.\eqref{eq:exterior derivative definition} twice, $d^2\alpha$ involves a sum over $k$-cells where each cell appears twice with opposite orientation and thus cancels. 

There is a natural inner product on the space of $k$-forms $\Omega^k$. Given an inner product $(\cdot, \cdot)$ on $A=\R, \Z ,\C$, then the inner product between $A$-valued $k$-forms is given by
\begin{equation}
    (\alpha, \omega)=\sum_{c_k} (\alpha(c_k), \omega(c_k)) \ ,
\end{equation}
i.e. we simply sum over their inner product on all $k$-cells. 

Another useful notion is that of the \textit{dual lattice}, which is again a standard construction in lattice field theory. This is another cell complex, where a $0$-cell (i.e. a dual site) is assigned to the center of each primal $n$-cell (where $n$ is the dimension). If two primal $n$-cells share an $(n-1)$-cell, then we assign a dual $1$-cell (i.e. a link) between the two corresponding dual sites, etc. For example, the dual of a hypercubic lattice is again hypercubic with the same lattice spacing but with all sites shifted by half a lattice spacing in all directions. In two dimensions, the honeycomb and triangular lattices are dual to each other. Call the space of dual lattice $k$-cells $\widetilde{C}_{k}$. Dual lattice $k$-forms are then just maps $\widetilde{C}_{k} \to A$. Cells in the primal and dual lattices are related by the star operator,
\begin{equation}
    \label{eq:star operator definition}
    \star :  C_k \leftrightarrow \widetilde{C}_{n-k} \ ,
\end{equation}
which in $n$ dimensions assigns to each primal $k$-cell the unique dual $(n-k)$-cell which intersects it, and vice versa. By convention, we choose $\star ^2=1$. The star operator also allows one to relate primal and dual lattice forms via the rule,
\begin{equation}
    \star\alpha(\widetilde{c}_{n-k})=\alpha(\star \widetilde{c}_{n-k}) \ .
\end{equation}
In other words, given a primal $k$-form $\alpha$, we can define a dual $(n-k)$-form $\star \alpha$ by assigning it a value on all dual lattice $(n-k)$ cells via the above rule. The star operator is an isometry, i.e. $(\star \alpha, \star \omega)=(\alpha,\omega)$.

The star operator is similar to the continuum Hodge star, but it differs in that our definition is purely combinatorial, while the continuum Hodge star depends on the metric; this is important especially on irregular lattices, to ensure that lattice quantities converge to their continuum counterpart. Some discussion of this can be found in \cite{Teixeira:2013kta}, but we will not discuss this further here. Via the star operator, one can define the co-derivative $\delta$ which acts on primal forms by lowering their degree,
\begin{equation}
    \label{eq:delta definition}
    \delta \equiv \star d \star : \Omega^k \to \Omega^{k-1} \ .
\end{equation}
Since $d^2=0$, then $\delta^2=0$. The key use of the co-derivative is in the integration by parts formula, which states that
\begin{equation}
    \label{eq:integration by parts formula}
    (d\alpha, \omega ) = (\alpha, \delta \omega) \ .
\end{equation}
It is easy to check that the form rank matches for the two forms on each side of this equation. Note that the definition of the star operator eq.\eqref{eq:star operator definition} hides a conventional choice for the orientation of dual lattice cells; here we choose the convention such that eq.\eqref{eq:integration by parts formula} holds as stated with $\delta$ defined as in eq.\eqref{eq:delta definition}. Note that this convention for the orientation of dual cells may not match an intuitive choice.

The lattice Laplacian is defined from the lattice derivatives as
\begin{equation}
    \label{eq:lattice laplacian}
    \square \equiv d \delta +\delta d  \ .
\end{equation}
With this choice, the lattice Laplacian is positive semi-definite, and on the hypercubic lattice it is given by the familiar expression (on a $0$-form $f$):
\begin{equation}
    \square f(x)=\delta d f(x) = \sum_{\mu } (2f(x)-f(x+\hat{\mu})-f(x-\hat{\mu})) \ .
\end{equation}
In Fourier space, this gives eq.\eqref{eq:lattice Laplacian momentum space}.

\subsection{Hodge decomposition}

In this section of the appendix, we state and prove the Hodge decomposition for $\R$-valued lattice differential forms. The restriction to $\R$ is due to the fact that one can see, say, $\Z$-valued forms as $\R$-valued forms, but the Hodge decomposition does not preserve their integer-valuedness. 

First note that the lattice Laplacian $\square$ defined in eq.\eqref{eq:lattice laplacian} does not change the form rank. A form which satisfies $\square \alpha=0$ is called \textit{harmonic}. These are exactly the zero modes of the Laplacian and therefore prevent its invertibility. Call the space of harmonic $k$-forms $\Omega_{\mathrm{harm}}^{k}$. Then since the space of $k$-forms is a finite-dimensional vector space, it can be split into the direct sum of the space of harmonic forms and its orthogonal complement according to the inner product $(\cdot, \cdot)$,
\begin{equation}
    \label{eq:direct sum decomp harmonic}
    \Omega^k =\Omega_{\mathrm{harm}}^{k} \oplus  \pqty{\Omega_{\mathrm{harm}}^{k}}^\perp \ .
\end{equation}
Note that the Laplacian is block-diagonal with respect to this decomposition, i.e. it takes the form $\square = \begin{pmatrix} 0 & 0 \\ 0 & \square \end{pmatrix}$ with a slight abuse of notation. This is because it maps forms in $\pqty{\Omega_{\mathrm{harm}}^{k}}^\perp$ to forms in $\pqty{\Omega_{\mathrm{harm}}^{k}}^\perp$. By definition, $\omega \in \pqty{\Omega_{\mathrm{harm}}^{k}}^\perp$ if it is orthogonal to all harmonic forms. But then given a generic harmonic form $\alpha$, integration by parts gives
\begin{equation}
    (\alpha, \square\omega)=(\square\alpha, \omega)=0 \ ,
\end{equation}
which means that $\square \omega$ is also orthogonal to all harmonic forms, and therefore belongs to $\pqty{\Omega_{\mathrm{harm}}^{k}}^\perp$. This also means that the Laplacian $\square$ is invertible when restricted to $\pqty{\Omega_{\mathrm{harm}}^{k}}^\perp$ (since there are no zero modes there).

The lattice Green's function is defined as the pseudo-inverse of $\square$, i.e.
\begin{equation}
    G(\alpha) = \begin{cases} 0 & \alpha \in \Omega_{\mathrm{harm}}^{k}\\
    (\square)^{-1}\alpha & \alpha \in \pqty{\Omega_{\mathrm{harm}}^{k}}^\perp
    \end{cases}
\end{equation}
Now note the following fact. Let $\Omega = \oplus_k \Omega^k$ be the vector space of all lattice $k$-forms. Suppose that $L: \Omega \to \Omega$ is a linear operator which commutes with the Laplacian $\square$. Then it commutes with $G$. Note that if the Laplacian had been invertible, this result would have been trivial. In the general case, the result follows by writing $L$ in block matrix form, imposing that it commutes with $\square$ and using the invertibility of $\square$ in $\pqty{\Omega_{\mathrm{harm}}^{k}}^\perp$.

It is not hard to see that $d, \delta$ commute with the Laplacian. Therefore, they also commute with $G$. Thus every form in $\pqty{\Omega_{\mathrm{harm}}^{k}}^\perp$ may be written as
\begin{equation}
    \alpha = \square^{-1} (d\delta +\delta d) \alpha = d \square^{-1} \delta \alpha + \delta \square^{-1} d \alpha \ .
\end{equation}
This is then the Hodge decomposition, i.e. given an arbitrary $k$-form $\alpha$ there exist a $(k-1)$-form $\alpha_1$, a $(k+1)$-form $\alpha_2$ and a harmonic $k$-form $\alpha_3$ such that 
\begin{equation}
    \alpha = d \alpha_1 + \delta \alpha_2 + \alpha_3 \ .
\end{equation}
Note that if $\alpha$ is, say, integer-valued, generally this is not going to be the case for $\alpha_1,\alpha_2$ since they involve $\square^{-1}$. Note further that the three forms in the decomposition are orthogonal. For the first two, this follows from integrating by parts,
\begin{equation}
    (d\alpha_1, \delta \alpha_2 )= (  d^2 \alpha,  \alpha) = 0 \ .
\end{equation}
For $\alpha_3$, note that for any harmonic form $\alpha$, integration by parts gives
\begin{equation}
    0=(\alpha, \square \alpha)=(d\alpha, d\alpha) + (\delta\alpha, \delta \alpha) \ .
\end{equation}
Since the inner product is positive definite, $d\alpha=\delta\alpha=0$ and therefore by a similar argument $\alpha_3$ is orthogonal to $d\alpha_1, \delta\alpha_2$. Orthogonality also implies that one can replace integration over $\alpha$ in a path-integral by integration over $\alpha_1, \alpha_2, \alpha_3$ up to an overall constant.

In infinite volume, such as in the discussion in Section \ref{sec:monopole gas}, one implicitly imposes boundary conditions such that forms go to zero at infinity. With this requirement, there are no harmonic forms and the harmonic term drops out of the Hodge decomposition (as in eq.\eqref{eq:hodge decomp main text}). In finite volume, one has to include harmonic forms which contribute \textquote{winding} to the partition function. These contributions are typically exponentially decaying with the lattice size and can usually (but not always) be neglected.

\subsection{Lattice Poincaré lemma}

Here we give a brief discussion of the lattice Poincaré lemma, which we used a couple of times in the main text. The general problem solved by the lattice Poincaré lemma is finding the general solution to the equation $d \alpha=0$ for a generic $k$-form $\alpha$. This equation appears generically when performing duality transformations on the lattice, and is generally the cause of the difference between infinite-volume and finite-volume dualities. For a constructive proof of the Poincaré lemma on the lattice either in infinite volume or with periodic boundary conditions, see \cite{Lscher1999, Kadoh:2003ii}.

Forms $\alpha$ which satisfy $d \alpha=0$ are known as \textit{closed}. Denote the space of closed $k$-forms as $Z^k$. Forms $\alpha$ which can be written as $\alpha=d\omega$ for some other form $\omega$ are known as \textit{exact}. Denote the space of exact $k$-forms as $B^k$. Since $d^2=0$ it is clear that every exact form is closed, $B^k \subset Z^k$. The failure of a closed form to be exact is captured by the $k$-th \textit{cohomology group}, defined as the quotient of the space of closed forms by that of exact forms,
\begin{equation}
    H^k = \frac{Z^k}{B^k} \ .
\end{equation}
An element $[\alpha] \in H^k$ is an equivalence class of closed forms where two closed forms $\alpha, \alpha'$ are in the same class, $[\alpha]=[\alpha']$, if and only if they differ by an exact form, $\alpha-\alpha'=d\omega$ for some $\omega$. Concretely, this means that we can choose representatives $\alpha_i$ for each cohomology class (i.e. for each element of $H^k$) and every closed $k$-form $\alpha$ can be written as
\begin{equation}
    \alpha = d\omega + \alpha_i \ ,
\end{equation}
for some $\alpha_i$ and some $\omega$. In lattice field theory calculations, the typical application of this result is in removing a constraint from a partition function. Suppose $\alpha$ is a $k$-form. Then the above result implies that for an arbitrary function $f$,
\begin{multline}
    \pqty{\prod_{c_k} \int_A D\alpha(c_k) } \delta(d\alpha) f(\alpha) =\\=\mathcal{N} \sum_{\alpha_i \in H^k}\pqty{\prod_{c_{k-1}} \int_A D\omega(c_{k-1}) } f(d\omega + \alpha_i) \ .
\end{multline}
Note that we used capital $D$ for the integration measure $D\alpha(c_k)$ over all values $\alpha(c_k) \in A$ to avoid confusion with the exterior derivative. In other words, we can remove the closedness constraint from $\alpha$ by summing over all exact forms up to contributions indexed by $\alpha_i$. In the above formula $\mathcal{N}$ is a (possibly infinite) normalization constant, which cancels on all expectation values. Note that, for example for $\R$, the sum over $H^k$ may in fact be an integral. If $H^k=0$, then every closed form is exact and we can remove $\alpha_i$. An introduction to cohomology and how to compute it can be found in the previously cited reference \cite{nakahara2003geometry}. If the lattice provides a cell decomposition of a manifold, then the cohomology groups of lattice differential forms coincide with the continuum cohomology groups of the underlying manifold. Two cases of most direct interest to lattice field theory are discretizations of $\R^n$ (i.e. \textquote{infinite volume}) and the torus $(S^1)^n$. In more complicated case, such as a torus with twisted or open boundary conditions, similar results can be derived (see the appendix in \cite{hornung2021mass} for a specific example) but they must be done on a case by case basis, and in particular statements such as integration by parts may also fail.

The cohomology groups also depend on the underlying Abelian group $A$ and the dependence on this and the underlying manifold $M$ is denoted as $H^k(M;A)$. In particular, one has
\begin{equation}
    \label{eq:R^n cohomology}
    H^k(\R^n;\R) = \begin{cases} \R & k=0\\
        0  & \mathrm{otherwise}
        \end{cases} \ .
\end{equation}
This reflects the fact that, in infinite volume, closed zero forms are just constant and this constant is an arbitrary real value; on the other hand, all closed $k$-forms with $k \geq 1$ are exact. We used this fact to obtain eq.\eqref{eq:m is exact} in the main text. If we had taken $A=\Z$, then the relevant group would have been the same up to the replacement $\R \to \Z$ on the right-hand side of eq.\eqref{eq:R^n cohomology}. 

On a torus, instead, one has non-trivial topology due to non-contractible cycles. An example may help visualize why they're important. Suppose that we have a $1$-form $\alpha$, and we want to show that $\alpha=d\omega$ for some $\omega$. Then a necessary and sufficient condition is that given an arbitrary closed path $\gamma$ on the lattice,
\begin{equation}
    \label{eq:zero on all closed paths}
    \sum_{c_1 \in \gamma} \alpha(c_1)=0 \ .
\end{equation}
It is clear that if $\alpha$ is exact, i.e. $\alpha_{l}=\omega_y-\omega_x$ at link $l:x\to y$, then successive terms in the sum appear with opposite sign, thus canceling and leading to a result equal to zero. As we will show in just a moment, the converse is also true. But note that if $\gamma$ is the boundary of a plaquette, then eq.\eqref{eq:zero on all closed paths} is just $d\alpha$ which equals zero due to the closedness of $\alpha$. All contractible loops can be built by tiling them with plaquettes and therefore the fact that $\alpha$ is closed implies that eq.\eqref{eq:zero on all closed paths} is satisfied on all contractible lattice loops. The non-contractible loops are then precisely the obstruction to exactness of closed forms.

Now suppose that eq.\eqref{eq:zero on all closed paths} holds with $\alpha$ closed; we want to show that $\alpha$ is exact. We define $\omega_x$ explicitly by choosing an arbitrary starting point $x_0$ and an arbitrary path (i.e. a sequence of oriented links) $\gamma:x_0 \to x$ and set,
\begin{equation}
    \label{eq:omega definition}
    \omega_x = \sum_{c_1 \in \gamma} \alpha(c_1) \ .
\end{equation}
The choice of path and starting points are irrelevant as long as eq.\eqref{eq:zero on all closed paths} is satisfied, since changing path or starting point leads to a difference which equals the sum of $\alpha$ over a closed path, which is zero by assumption. Next, we show that indeed $d\omega=\alpha$. Choose a link $l:x \to y$ and define $\omega_x$ and $\omega_y$ according to eq.\eqref{eq:omega definition} by choosing an arbitrary path $\gamma$ for $\omega_x$; for $\omega_y$, instead, we first go from $x_0 \to x$ using $\gamma$ and then follow $l$ to reach $y$. Then all terms in $\omega_y-\omega_x$ cancel except the one at $l$, and we get $\omega_y-\omega_x=\alpha_l$.

The cohomology groups on the $n$-torus are given by
\begin{equation}
    H^k((S^1)^n ; \R)= \R^{{n \choose k}} \ .
\end{equation}
For $\Z$-valued forms, we can replace $\R$ with $\Z$ everywhere in the previous equation. For example, for the two-torus $H^1((S^1)^2; \R)=\R^2$, corresponding to non-contractible loops. This means that every closed $1$-form $\alpha$ on the torus can be written as
\begin{equation}
    \alpha = d\omega + \alpha_i \ ,
\end{equation}
where $\alpha_i$ contains the winding contributions along the two non-contractible loops. For $A=\R$, these contributions can be chosen to be harmonic forms by the Hodge decomposition. For example, on the square lattice  we can choose two arbitrary non-contractible loops $\gamma_1$ in the $x$ direction and $\gamma_2$ in the $y$ direction of length $L_i$ respectively, and compute the winding,
\begin{equation}
    w_i = \frac{1}{L_i}\sum_{c_1 \in \gamma_i} \alpha(c_1) \ .
\end{equation}
Then we can define $1$-forms $\alpha_i$ equal to $1$ on all links in direction $i$ and zero otherwise. Each $\alpha_i$ is a closed form and in fact harmonic. Then $w_1 \alpha_1 + w_2 \alpha_2$ has the same winding numbers as $\alpha$, meaning that $\alpha-\sum_i w_i \alpha_i$ is zero summed on every closed lattice path and therefore
\begin{equation}
    \alpha = d\omega +w_1 \alpha_1 + w_2 \alpha_2 \ ,
\end{equation}
for a form $\omega$ constructed as previously discussed. In the main text, we have ignored the sum over winding contributions.

For integer-valued forms, since the Hodge decomposition does not map integer-valued forms to integer-valued forms, one cannot choose the representatives of cohomology classes to be harmonic. Nonetheless the same result applies, with a different choice of representative. In particular, we would compute the winding number without dividing,
\begin{equation}
    w_i = \sum_{c_1 \in \gamma_i} \alpha(c_1) \ .
\end{equation}
Thus if $\alpha \in \Z$, then $w_i \in \Z$. Then  we choose $\alpha_i$ equal to one on a set of links intersecting a non-contractible dual  lattice loop in the opposite direction (in higher dimension a closed non-contractible $(n-1)$-dimensional dual hypersurface), so that $\alpha_i$ has winding $1$ for every non-contractible loop in the $i$ direction. Then $w_1 \alpha_1 + w_2 \alpha_2$ is closed (but not harmonic) and has the same winding as $\alpha$, leading to the same final result. 

\end{document}